\documentclass[11pt]{article}
\usepackage{amsmath}
\usepackage{latexsym}
\usepackage{amsfonts}
\usepackage{amssymb}
\newtheorem{definition}{Definition}

\newtheorem{proposition}{Proposition}
\newcommand{\erf}{{\rm{erf}}}
\newcommand{\Ai}{{\rm{Ai}}}
\newcommand{\eq}{\label}
\newcommand{\al}{\alpha}
\newcommand{\be}{\beta}
\newcommand{\ga}{\gamma}
\newcommand{\de}{\delta}
\newcommand{\ep}{\epsilon}
\newcommand{\half}{{\lower0.25ex\hbox{\raise.6ex\hbox{\the\scriptfont0 1}\kern-.2em\slash\kern-.1em\lower.25ex\hbox{\the\scriptfont0 2}}}}
\newcommand{\thalf}{{\lower0.25ex\hbox{\raise.6ex\hbox{\the\scriptfont0 3}\kern-.2em\slash\kern-.1em\lower.25ex\hbox{\the\scriptfont0 2}}}}
\newcommand{\fhalf}{{\lower0.25ex\hbox{\raise.6ex\hbox{\the\scriptfont0 5}\kern-.2em\slash\kern-.1em\lower.25ex\hbox{\the\scriptfont0 2}}}}
\newcommand{\othird}{{\lower0.25ex\hbox{\raise.6ex\hbox{\the\scriptfont0 1}\kern-.2em/\kern-.1em\lower.25ex\hbox{\the\scriptfont0 3}}}}

\begin{document}
\title{
% \begin{flushright}
%   \small MS-99/?
% \end{flushright}
% \vskip1.0cm
\large\bf
Gap Probabilities for Edge Intervals in Finite Gaussian and Jacobi
Unitary Matrix Ensembles}
\author
{\large N.S. Witte,\\
{\it Department of Mathematics and Statistics}\\
{\&} {\it School of Physics, University of Melbourne}\\ 
{\it Parkville, Victoria 3052, AUSTRALIA}\\
{\large P.J. Forrester},\\
{\it Department of Mathematics and Statistics}\\
{\it University of Melbourne}\\
{\it Parkville, Victoria 3052, AUSTRALIA}\\
and {\large Christopher M. Cosgrove}\\
{\it School of Mathematics and Statistics}\\
{\it University of Sydney}\\
{\it Sydney, NSW 2006, AUSTRALIA.} }
\maketitle
\begin{abstract}
The probabilities for gaps in the eigenvalue spectrum of the finite
dimension $ N \times N $ random matrix Hermite and Jacobi unitary ensembles 
on some single and disconnected double intervals are found.
These are cases where a reflection symmetry exists and the probability factors
into two other related probabilities, defined on single intervals. 
Our investigation uses the system of partial differential equations 
arising from the Fredholm determinant expression for the gap probability
and the differential-recurrence equations satisfied by Hermite and Jacobi
orthogonal polynomials.
In our study we find second and third order nonlinear ordinary differential
equations defining the probabilities in the general $N$ case. For $N=1$
and $N=2$ the probabilities and thus the solution of the equations are
given explicitly. An asymptotic expansion for large gap size is obtained from 
the equation in the Hermite case, and also studied is the scaling at the 
edge of the Hermite spectrum as $ N \to \infty $, and the Jacobi to Hermite
limit; these last two studies make correspondence to other cases reported
here or known previously. 
Moreover, the differential equation arising in the Hermite ensemble is
solved in terms of an explicit rational function of a \mbox{Painlev\'e-V}
transcendent and its derivative, and an analogous solution is provided
in the two Jacobi cases but this time involving a \mbox{Painlev\'e-VI}
transcendent.
\end{abstract}
\medskip
\noindent
AMS classification - primary: 15A52, secondary: 34A34, 34A05, 33C45
\vfill\eject
\section{Introduction}
It is a celebrated discovery of Jimbo, Miwa, M\^ori and Sato
\cite{JMMS-80} that the probability of an eigenvalue free region in
the bulk of the infinite GUE (Gaussian unitary ensemble of random Hermitian 
matrices) can be expressed exactly in terms of a \mbox{Painlev\'e-V} 
transcendent. This has initiated a number of studies 
\cite{TW-94, ASV-95, HS-99} which, in addition to clarifying the general setting
of the exact result, give formalisms that allow analogous results to be
obtained in other cases. For example, the probability of an eigenvalue
free region at the edge of the infinite GUE (appropriately scaled) has
been expressed in terms of the \mbox{Painlev\'e-II} transcendent
\cite{TW-94a}, while a \mbox{Painlev\'e-III} transcendent has been shown to
determine the distribution of the smallest eigenvalue in the infinite
LUE (Laguerre unitary ensemble of non-negative matrices of the form 
$A^\dagger A$ with $A$ complex \cite{TW-94}).
Moreover, for the finite classical ensembles with unitary symmetry, the
probability of a single eigenvalue free region which includes an
endpoint of the support of the weight has been expressed in terms of 
solutions of certain non-linear equations \cite{TW-94}.

It is the objective of this work to provide three new evaluations of
gap probabilities for particular finite classical random matrix
ensembles with unitary symmetry. The ensembles considered are the
finite GUE and the symmetric Jacobi unitary ensemble (JUE). We recall
the eigenvalue p.d.f.~for an ensemble with unitary symmetry is of the
form
\begin{equation}
  \prod_{l=1}^N w_2(\lambda_l) \prod_{1 \le j < k \le N}
  |\lambda_k - \lambda_j |^2,
\label{pdf}
\end{equation}
where the weight function $w_2(\lambda)$ determines the specific unitary
ensemble:
\begin{equation}
  w_2(\lambda) = \left \{ \begin{array}{ll}
  e^{-\lambda^2}, & {\rm GUE} \\
  (1-\lambda)^{\alpha}(1+\lambda)^{\beta}, & {\rm JUE} \end{array} \right.
\label{weights}
\end{equation}
(the symmetric Jacobi ensemble refers to the case $\alpha=\beta$ of the JUE).
These ensembles can be realised in terms of matrices with independent
Gaussian elements. For the GUE, each matrix $X$ say must be Hermitian and
have its diagonal elements $x_{jj}$ (which must be real) 
and upper triangular elements $x_{jk} = u_{jk} + i v_{jk}$ chosen with
p.d.f.~N$(0,1/\sqrt{2})$ and N$(0,1/2) + i {\rm N}(0,1/2)$. For the JUE
one first constructs \cite{Mu-82} auxiliary rectangular
$M_1 \times N$ and $M_2 \times N$ ($M_1, M_2 \ge N$) matrices
$a$ and $b$ respectively, with complex elements independently distributed
according to N$(0,1/\sqrt{2}) + i {\rm N}(0,1/\sqrt{2})$. Then with
$A = a^\dagger a$, $B= b^\dagger b$, it can be shown that the
distribution of the eigenvalues of $A(A+B)^{-1}$ is an example of
(\ref{pdf}) with
\begin{equation}
  w_2(\lambda) = \lambda^{M_1 - N}(1 - \lambda)^{M_2 - N}, \quad
  0 < \lambda < 1.
\label{jue}
\end{equation}
The change of variables $\lambda \mapsto (1 - \lambda)/2$ shows that this
is an example of the JUE.

With $E_{2}(0;I;w_2(\lambda);N)$ denoting the probability that there are no
eigenvalues in the interval $I$ of an ensemble with eigenvalue
p.d.f.~(\ref{pdf}), the probabilities to be calculated are
\begin{equation}
  E_{2}(0;(-\infty,-s) \cup (s,\infty);e^{-\lambda^2};N)
\label{E-gue}
\end{equation}
and
\begin{equation}
  E_{2}(0;(-1,-s) \cup (s,1); (1-\lambda^2)^{\alpha};N), \qquad
  E_{2}(0;(-s,s);(1-\lambda^2)^{\alpha};N).
\label{E-jue}
\end{equation}
The quantity (\ref{E-gue}) gives the probability that there are no eigenvalues
in the GUE with modulus greater than $s$, as does the first quantity in
(\ref{E-jue}) for the JUE. The final quantity in (\ref{E-jue}) gives the 
probability that there are no eigenvalues in the JUE with modulus less than
$s$ (the analogous probability for the GUE has previously been
evaluated \cite{TW-94}).

These probabilities are special because, like the situation in which
$I$ is a single interval which includes an endpoint of the support of
the weight function noted above, we will show that they can be evaluated
exactly in terms of the solution of certain non-linear equations.
Another motivating factor for our study is a recently discovered
\cite{Fo-99} general identity satisfied by the probability
$E_2(0;I;w_2(\lambda);N)$ applicable whenever $w_2(\lambda)$ is even and
$I$ is symmetrical about the origin. The identity states
\begin{align}
  E_2(0;I;w_2(\lambda);N) = 
  & E_2(0;I^+;y^{-1/2} w_2(y^{1/2});[(N+1)/2]) \nonumber \\*
  & \times E_2(0;I^+;y^{+1/2} w_2(y^{1/2});[N/2])
\label{E-split}
\end{align}
where $I^+$ denotes the portion of $I$ on ${\bf R}^+$ in the variable
$y=\lambda^2$ and $w_2(y^{1/2}) := 0$ for $y < 0$. The probabilities
(\ref{E-gue}) and (\ref{E-jue}) are all of the form required by the LHS of this
identity, and are thereby related to
\begin{equation}
  E_2(0;(s^2,\infty);y^{\mp 1/2} e^{-y};m)
\label{E-laguerre}
\end{equation}
and
\begin{equation}
  E_{2}(0;(s^2,1);y^{\mp 1/2} (1-y)^{\alpha};m), \quad
  E_{2}(0;(0,s^2);y^{\mp 1/2} (1-y)^{\alpha};m)
\label{E-jacobi}
\end{equation}
($m = [(N+1)/2], [N/2]$) respectively. The weight functions in 
(\ref{E-laguerre}) and (\ref{E-jacobi}) are again classical, 
but now the eigenvalues are excluded from
a single interval which includes an endpoint of the support of the weight
function. The corresponding probabilities are known in terms of
\mbox{Painlev\'e-V} and VI transcendents from the works of \cite{TW-94} and
\cite{HS-99}. Thus as a by-product of providing independent evaluations of the
LHS of (\ref{E-split}), we will be providing inter-relationships between
solutions of nonlinear equations. This theme is further developed in
\cite{Fo-99}.

In Section 2 the formalism of Tracy and Widom \cite{TW-94} giving coupled 
differential equations for the gap probability and some auxiliary quantities 
is revised. In Section 3 these coupled equations are reduced to a single 
ordinary differential equation specifying the probability (1.4), and this 
equation is used to compute the large gap size behaviour of the probability, 
as well as various scaling limits. A similar study is undertaken in Section 
4 for the probabilities (1.5). In Section 5 the solution of the second order
equations of Section 3 (Hermite case) and third order equations of Section 4 
(Jacobi case) are given in terms of certain \mbox{Painlev\'e-V} and 
\mbox{Painlev\'e-VI} transcendents respectively.

\vfill\eject
\setcounter{equation}{0}

\section{The General Formalism}
We are going to consider a set of the eigenvalue spectrum arising from the 
matrix ensembles consisting of an arbitrary number of disconnected intervals.
In general the eigenvalues may be excluded from $ M $ disjoint intervals which
together form $ I $. Thus with the endpoints of these intervals denoted
$ \{a_{j}\}^{2M}_{j=1} $,
\begin{equation}
  I = \bigcup^{M}_{m \geq 1} (a_{2m-1},a_{2m}) \ .
\label{interval}
\end{equation}
The probability of no eigenvalues being found in this interval is given by the
general expression (see e.g. \cite{F-00}) 
\begin{equation}
   E(0;I) = 1 + \sum^{\infty}_{n=1} {(-1)^n \over n!}
    \int_{I} dx_{1} \ldots \int_{I} dx_{n}
    \rho_{n}(x_{1},\ldots ,x_{n}) \ ,
\label{prob-0}
\end{equation}
where $ \rho_{n} $ is the $ n $-point distribution function of the eigenvalue
p.d.f. For matrix ensembles with unitary symmetry the eigenvalue p.d.f. is
proportional to (\ref{weights}), and the corresponding $ n $-point 
distribution function is given in terms of the orthonormal polynomials 
$ \{p_{j}(x)\}_{j = 0,1,2,\ldots} $ associated with the weight function
$ w_2(x) $ according to the formula
\begin{equation}
  \rho_{n}(x_{1},\ldots ,x_{n}) = 
    \det \left[ K_N(x_{i},x_{j}) \right]_{1\leq i,j \leq n} \ ,
\label{n-particle}
\end{equation}
where 
\begin{equation}
  K_{N}(x,y) = \left[ w_2(x)w_2(y) \right]^{1/2}
  \sum^{N-1}_{l=0} p_{l}(x)p_{l}(y) \ .
\label{kernel-sum}
\end{equation}
Substituting into (\ref{prob-0}), the Fredholm theory of integral operators
then gives
\begin{equation}
  E(0;I) = \det( \Bbb{I}-\Bbb{K}_{N} ) \ ,
\label{prob-fd}
\end{equation}
where $ \Bbb{K}_{N} $ is the integral operator with kernel $ K_{N}(x,y) $ 
defined on the interval $ I $. A crucial point is that (\ref{kernel-sum}) can
be summed according to the Christoffel-Darboux formula and so written in the
special form \cite{IIKS-90}
\begin{equation}
   K_{N}(x,y) = { \phi(x)\psi(y) - \phi(y)\psi(x) \over x-y } \ ,
\label{kernel}
\end{equation}
where with $ a_{N} $ denoting the coefficient of $ x^N $ in $ p_{N}(x) $
\begin{equation}
 \begin{split}
    \phi(x) & = 
    \left( {a_{N-1} \over a_{N}}w_2(x) \right)^{1/2} p_{N}(x) \ ,
    \\
    \psi(x) & = 
    \left( {a_{N-1} \over a_{N}}w_2(x) \right)^{1/2} p_{N-1}(x) \ .
 \end{split}
\label{phi-psi}
\end{equation}
We suppose furthermore that $ \phi(x) $ and $ \psi(x) $ satisfy the 
recurrence-differential relations
\begin{equation}
 \begin{split}
  m(x)\phi'(x)
  & = A(x)\phi(x)+B(x)\psi(x) \ , \\
  m(x)\psi'(x)
  & = -C(x)\phi(x)-A(x)\psi(x) \ ,
\end{split}
\label{rde}
\end{equation}
where the coefficient functions $ m(x), A(x), B(x), C(x) $ are polynomials in $ x $.

In the above setting Tracy and Widom \cite{TW-94} have derived a set of
coupled differential equations for the determinant (\ref{prob-fd}), as well as 
some auxiliary quantities. The latter we introduce in the following definitions.
\begin{definition}
Let $ A \doteq A(x,y) $ denote that the integral operator $ A $ has kernel
$ A(x,y) $. Then the kernels $ \rho(x,y) $ and $ R(x,y) $ are specified by
\begin{equation}
  \begin{split}
     (1-K)^{-1} 
   & \doteq \rho(x,y) \ ,
   \\
     K(1-K)^{-1} 
   & \doteq R(x,y) \ ,
  \end{split}
\label{resolvent}
\end{equation}
\end{definition}
The operator $ K(1-K)^{-1} $ is called the resolvent and $ R(x,y) $ the 
resolvent kernel.
\begin{definition}
For $ k \in \Bbb{Z}_{\geq 0} $ the functions $ Q_{k} $ and $ P_{k} $ are 
defined by
\begin{equation}
  \begin{split}
     Q_{k}(x)  
   & = \int_{I}dy\; \rho(x,y) y^{k}\phi(y) \ ,
   \\
     P_{k}(x)
   & = \int_{I}dy\; \rho(x,y) y^{k}\psi(y) \ ,
  \end{split}
\label{QP-defn}
\end{equation}
and their values at the endpoints $ a_{j} $ of $ I $ are denoted 
$ q_{kj}, p_{kj} $ so that 
\begin{equation}
  \begin{split}
     q_{kj}
   & = Q_{k}(a_{j}) \equiv \lim_{x \to a_{j}} Q_{k}(x)
   \\
     p_{kj}
   & = P_{k}(a_{j}) \equiv \lim_{x \to a_{j}} P_{k}(x)
  \end{split}
\label{qp-defn}
\end{equation}
\end{definition}
Where there is no confusion, we denote $ q_{0j}, p_{0j} $ by $ q_{j}, p_{j} $.
\begin{definition}
The inner products $ u, v, w $ are defined by
\begin{equation}
  \begin{split}
     u
   & = \langle \phi | Q \rangle = \int_{I}dy\; Q_{0}(y)\phi(y) \ ,
   \\
     v
   & = \langle \psi | Q \rangle = \int_{I}dy\; Q_{0}(y)\psi(y)
     = \langle \phi | P \rangle = \int_{I}dy\; P_{0}(y)\phi(y) \ ,
   \\
     w
   & = \langle \psi | P \rangle = \int_{I}dy\; P_{0}(y)\psi(y) \ .
  \end{split}
\label{uvw-defn}
\end{equation}
\end{definition}

Our goal is to characterise the probabilities (\ref{E-gue}) and (\ref{E-jue})
as the solution of certain non-linear differential equations. Following
\cite{TW-94} this is achieved by specifying partial differential equations for
the quantities $ q_{j}, p_{j}, u, v, w $ and $ R(a_{j},a_{k}) $.
These equations come in two types: a set of universal equations which are
independent of the recurrence-differential equations (\ref{rde}), and a 
second set of equations which depend on the details of (\ref{rde}).
Let us first present the former.
\begin{proposition}
For general functions $ \phi(x),\psi(x) $ we have the relations
\begin{equation}
  {\partial \over \partial a_{j}} \log \det (1-K)
  = (-1)^{j-1} R(a_{j},a_{j}) \ ,
\label{diff-E}
\end{equation}
and for $  j \neq k $,
\begin{equation}
  R(a_{j},a_{k})
  = {q_{j}p_{k} - p_{j}q_{k} \over a_{j}-a_{k}} \ ,
\label{fn-R}
\end{equation}
and
\begin{equation}
  {\partial \over \partial a_{k}} R(a_{j},a_{j})
  = (-1)^{k}R(a_{j},a_{k})R(a_{k},a_{j}) \ ,
\label{diff-R}
\end{equation}
with
\begin{equation}
  \begin{split}
    {\partial q_{j} \over \partial a_{k}} 
   & = (-1)^{k} R(a_{j},a_{k}) q_{k}
   \ , \\
    {\partial p_{j} \over \partial a_{k}} 
   & = (-1)^{k} R(a_{j},a_{k}) p_{k}
   \ ,
  \end{split}
\label{diff-pq}
\end{equation}
and
\begin{equation}
  \begin{split}
    {\partial u \over \partial a_{k}}
   & = (-1)^{k} q^2_{k}
   \ , \\
    {\partial v \over \partial a_{k}}
   & = (-1)^{k} q_{k}p_{k}
   \ , \\
    {\partial w \over \partial a_{k}}
   & = (-1)^{k} p^2_{k}
   \ .
  \end{split}
\label{diff-uvw}
\end{equation}
\end{proposition}

The second set of equations, which depend on the details of (\ref{rde}),
give the $ j = k $ cases of (\ref{fn-R}) and (\ref{diff-pq}). Now for
weights (\ref{weights}) the equations hold for $ m(x) $ a quadratic 
and $ A(x), B(x), C(x) $ linear functions, and thus of the general form
\begin{equation}
  \begin{split}
  m(x) & = \mu_{0}+\mu_{1}x+\mu_{2}x^2 \ ,
  \\
  A(x) & = \alpha_{0}+\alpha_{1}x \ ,
  \\
  B(x) & = \beta_{0}+\beta_{1}x  \ ,
  \\
  C(x) & = \gamma_{0}+\gamma_{1}x \ .
  \end{split}
\label{poly-defn}
\end{equation}
One then has the following equations \cite{TW-94}.
\begin{proposition}
In the case that $ \phi(x), \psi(x) $ satisfy the equations (\ref{rde})
with coefficient functions (\ref{poly-defn}) we have
\begin{equation}
  \begin{split}
    m_{i}{\partial q_{i} \over \partial a_{i}}
   & = [\alpha_{0}+\alpha_{1}a_{i}+\gamma_{1}u-\beta_{1}w-\mu_{2}v] q_{i}
   \\
   & \qquad
     + [\beta_{0}+\beta_{1}a_{i}+2\alpha_{1}u+2\beta_{1}v+\mu_{2}u] p_{i}
   \\
   & \qquad\qquad
     - \sum^{2M}_{k \neq i} (-1)^k R(a_{i},a_{k}) q_{k}m_{k}
   \ , \\
    m_{i}{\partial p_{i} \over \partial a_{i}}
   & = [-\gamma_{0}-\gamma_{1}a_{i}+2\gamma_{1}v+2\alpha_{1}w-\mu_{2}w] q_{i}
   \\
   & \qquad
     + [-\alpha_{0}-\alpha_{1}a_{i}+\beta_{1}w-\gamma_{1}u+\mu_{2}v] p_{i}
   \\
   & \qquad\qquad
     - \sum^{2M}_{k \neq i} (-1)^k R(a_{i},a_{k}) p_{k}m_{k}
   \ , \\
  \end{split}
\label{diff-diag-pq}
\end{equation}
and
\begin{align}
    m_{i}R(a_{i},a_{i})
   & = [\gamma_{0}+\gamma_{1}a_{i}-2\gamma_{1}v-2\alpha_{1}w+\mu_{2}w] q^2_{i}
   \nonumber \\*
   & \qquad
     + [\beta_{0}+\beta_{1}a_{i}+2\alpha_{1}u+2\beta_{1}v+\mu_{2}u] p^2_{i}
   \nonumber \\*
   & \qquad\qquad
     + [\alpha_{0}+\alpha_{1}a_{i}+\gamma_{1}u-\beta_{1}w-\mu_{2}v] 2q_{i}p_{i}
   \nonumber \\*
   & \qquad\qquad\qquad
     + \sum^{2M}_{k \neq i} (-1)^k m_{k} 
       { [q_{i}p_{k}-p_{i}q_{k}]^2 \over a_{i}-a_{k} } \ ,
\label{diag-R}
\end{align}
and furthermore,
\begin{align}
    {\partial \over \partial a_{i}}\left[ m_{i}R(a_{i},a_{i}) \right]
   & = 2\alpha_{1} q_{i}p_{i} + \beta_{1} p^2_{i} + \gamma_{1} q^2_{i}
   \nonumber \\*
   & \qquad\qquad
     - \sum^{2M}_{k \neq i} (-1)^k m_{k} R^2(a_{i},a_{k}) \ ,
\label{diff-diag-R}
\end{align}
where $ m_{i} = m(a_{i}) $.
\end{proposition}

\vfill\eject
\setcounter{equation}{0}

\section{The Gaussian Ensemble}

In this section we take up the problem of computing (\ref{E-jue}). We recall
(see e.g. \cite{ops-S}) that for the Gaussian weight $ e^{-x^2} $ the 
corresponding orthonormal polynomials $ \{p_{N}(x)\}_{N=0,1,\ldots} $ are given
in terms of the standard Hermite polynomials $ H_{N}(x) $ by
\begin{equation}
  p_{N}(x) = 2^{-N/2}\pi^{-1/4}(N!)^{-1/2} H_{N}(x) \ ,
\label{hermite-poly}
\end{equation}
and for the coefficient $ a_{N} $ of $ x^N $ in $ p_{N}(x) $ we have
\begin{equation}
  a_{N} = 2^{N/2}\pi^{-1/4}(N!)^{-1/2} \ .
\label{hermite-coeff}
\end{equation}
Substituting in (\ref{phi-psi},\ref{rde}) and making use of the differential 
relation for the Hermite polynomials gives \cite{TW-94}
\begin{equation}
 \begin{split}
 \phi'(x)
  & = -x\phi(x)+\sqrt{2N}\psi(x) \ , \\
 \psi'(x)
  & = +x\psi(x)-\sqrt{2N}\phi(x) \ , \\
\end{split}
\label{gauss-rde}
\end{equation}
and so we have
\begin{equation}
  m(x) = 1, \quad \alpha_1 = -1, \quad \beta_0 = \gamma_0 = \sqrt{2N},
  \quad \alpha_0 = \beta_1 = \gamma_1 = 0 \ .
\label{gauss-param}
\end{equation}

\subsection{Differential Equation}
The interval from which the eigenvalues are excluded in the probability
(\ref{E-gue}) consists of the union of the two intervals $ (-\infty,-s) $
and $ (s,\infty) $. Thus in (\ref{interval}) we have $ M = 2 $, and 
$ a_1 = -\infty $, $ a_2 = -s $, $ a_3 = s $ and $ a_4 = \infty $.
Because $ a_1 $ and $ a_4 $ are infinite, only the quantities in Propositions
1 and 2 relating to $ a_2 $ and $ a_3 $ are of interest. Furthermore it turns
out that the evenness symmetry of $ I $ and the odd/even symmetry of 
(\ref{hermite-poly}) implies the equations relating to $ a_2 $ are equivalent
to those relating to $ a_3 $. Thus $ \phi(-x) = (-1)^{N}\phi(x) $,
$ \psi(-x) = (-1)^{N-1}\psi(x) $, and so 
$ K_{N}(-x,-y) = K_{N}(x,y) $, which in turn implies 
$ \rho(-x,-y) = \rho(x,y) $ and so $ q_2 = (-1)^{N}q_3 $, 
$ p_2 = (-1)^{N-1}p_3 $, $ R(-x,-y) = R(x,y) $.
\begin{proposition}
The coupled differential equations for $ q = q_3, p = p_3 $ and 
$ R(s), R(-s,s) $ of the finite $ N $ GUE on the interval 
$ I = (-\infty,-s)\cup (s,\infty) $ are
\begin{align}
 {d \over ds} \ln E_{2}
  & = 2R(s) \ , 
  \label{gauss-sde:a}\\
  q'
  & = -sq+p[\beta_0-2u]+2q^2p/s \ , 
  \label{gauss-sde:b}\\
  p'
  & = +sp-q[\gamma_0+2w]-2qp^2/s \ , 
  \label{gauss-sde:c}\\
  u'
  & = -2q^2 \ , 
  \label{gauss-sde:d}\\
  w'
  & = -2p^2 \ , 
  \label{gauss-sde:e}\\
  R(-s,s)
  & = (-1)^{N-1}qp/s \ , 
  \label{gauss-sde:f}\\
  R(s)
  & = q^2[\gamma_0+2w]+p^2[\beta_0-2u]-2sqp+2q^2p^2/s \ , 
  \label{gauss-sde:g}\\
  R'(s)
  & = -2qp-2q^2p^2/s^2 \ ,
  \label{gauss-sde:h}
\end{align}
where primes indicate derivatives with respect to $ s $.
\end{proposition}
Proof - The first equation follows from (\ref{diff-E}), (\ref{gauss-sde:b})
and (\ref{gauss-sde:c}) follow from (\ref{diff-diag-pq}), 
(\ref{gauss-sde:d}) and (\ref{gauss-sde:e}) follow from (\ref{diff-uvw}),
(\ref{gauss-sde:f}) follows from (\ref{fn-R}), (\ref{gauss-sde:g}) from
(\ref{diag-R}) and (\ref{gauss-sde:h}) from (\ref{diff-diag-R}). In the
derivation of (\ref{gauss-sde:a}) and (\ref{gauss-sde:h}) use is made of
the general formula for the total derivative 
\begin{equation}
  {d \over ds}f_{I}(s,s) = \left.\left(
                           {\partial \over \partial a_{3}}f_{I}(a_{3},a_{3})
                          -{\partial \over \partial a_{2}}f_{I}(a_{3},a_{3})
                       \right)\right|_{a_{3}=-a_{2}=s}
  = 2{\partial \over \partial s}f_{I}(s,s) \ ,
\label{gauss-sde-proof}
\end{equation}
valid whenever $ f_{I}(a_2,a_2) = f_{I}(a_3,a_3) $.
$ \square $

There are a number of differences when one compares this set of coupled ODEs
with that arising from a single interval $ (s,\infty) $, as described in 
Ref.\cite{TW-94}. Firstly there is the appearance of factors of two, reflecting
the contributions from the two free endpoints, but more significantly is
the addition of nonlinear quadratic terms in (\ref{gauss-sde:b}, 
\ref{gauss-sde:c}, \ref{gauss-sde:g}, \ref{gauss-sde:h}). Regarding the 
reduction of these equations, it is straightforward to show that 
(\ref{gauss-sde:b}) and (\ref{gauss-sde:c}) together imply
\begin{equation}
  pq = uw-\half\beta_0 w+\half\gamma_0 u \ ,
\label{gauss-int1}
\end{equation}
or equivalently
\begin{equation}
  \beta_0\gamma_0 - 4pq = (\beta_0-2u)(\gamma_0+2w) \ ,
\label{gauss-int2}
\end{equation}
which is useful in the subsequent determination of a single equation for
$ R(s) $. In fact it is possible to reduce the coupled system of PDEs in
Proposition 3 down to a single ODE for $ R(s) $, or a single ODE for the
quantity
\begin{equation}
   \tilde{R}(s) = (-1)^N R(-s,s) \ .
\label{tilde-R}
\end{equation}

The boundary conditions satisfied by $ R(s) $ and $ \tilde{R}(s) $ are found 
from the large $ s $ form of the kernel 
\begin{equation}
  R(s) \underset{s \to \infty}{\sim} K_{N}(s,s) \ ,
\label{gauss-asympt}
\end{equation}
which implies 
\begin{equation}
 \begin{split}
 R(s)
  & \underset{s \to \infty}{\sim}
      {\pi^{-1/2}2^{-N} \over (N\!-\!1)!}e^{-s^2}
      \left[ H^2_{N}(s)-H_{N+1}(s)H_{N-1}(s) \right] \ , \\
 \tilde{R}(s)
  & \underset{s \to \infty}{\sim}
      -{\pi^{-1/2}2^{-N} \over (N\!-\!1)!} s^{-1} e^{-s^2}
             H_{N}(s)H_{N-1}(s) \ . \\
\end{split}
\label{gauss-bc}
\end{equation}
Of course we could replace the Hermite polynomials by their leading order
term, however of subsequent interest will be a scaling limit which requires
the full expression as written.

We can reduce this coupled system of PDEs into a single second order ODE,
and give two possible forms for this.
\begin{proposition}
The system of ODEs in Proposition 3 is equivalent to the following second-order
differential equation for $ \tilde{R}(s) $
\begin{align}
 &  \left[ s\tilde{R}'' + 2\tilde{R}' + 8Ns\tilde{R} + 24s^2\tilde{R}^2 
    \right]^2 = 
 \nonumber \\*
 &  \hspace{4.0cm}
    4[s+2\tilde{R}]^2
    \left[(\tilde{R}+s\tilde{R}')^2 + 8Ns^2\tilde{R}^2 + 16s^3\tilde{R}^3 
    \right] \ ,
\label{gauss-ode-b}
\end{align}
or to the second-order differential equation for $ R(s) $
\begin{equation}
  sR''+2R' = 2s[s-h] - 2h\sqrt{ (R+sR')^2 - 4s^2[s-h]R - 2Ns^2[s-h]^2} \ .
\label{gauss-ode}
\end{equation}
where we denote
\begin{equation}
  h \equiv \sqrt{s^2-2R'} \ .
\label{gauss-sqrtG}
\end{equation}
\end{proposition}
Proof - We adopt the notation $ a = sR $, $ b = s\tilde{R} $ for simplicity.
Using (\ref{gauss-sde:b}) and (\ref{gauss-sde:c}) together with the definition
(\ref{tilde-R}) we find the relation
\begin{equation}
  b' = q^2(\gamma_0+2w)-p^2(\beta_0-2u) \ ,
\end{equation}
and utilising (\ref{gauss-sde:g}) and (\ref{gauss-sde:h}) the corresponding
relation for $ a $ is
\begin{equation}
  a' - 4sb = q^2(\gamma_0+2w)+p^2(\beta_0-2u) \ .
\end{equation}
By squaring and subtracting the right hand sides of these two equations and
employing the integral (\ref{gauss-int2}) to eliminate the cross term we have
\begin{equation}
  (a')^2 = 8ab+8Nb^2+(b')^2 \ .
\label{gauss-red1}   
\end{equation}
There is also another relation which follows directly from (\ref{gauss-sde:h})
and reads
\begin{equation}
  sa' = a+2s^2b-2b^2 \ .
\label{gauss-red2}
\end{equation}
If one proceeds to eliminate $ a $ using (\ref{gauss-red1}) and 
(\ref{gauss-red2}) then a second-order differential equation is obtained for 
$ \tilde{R} $, namely (\ref{gauss-ode-b}). However if one eliminates $ b $ 
then a more useful second order equation for $ R $ emerges
\begin{multline}
 \half\left\{ s[s\mp\sqrt{s^2-2R'}] - \half a'' \right\}^2 = \\
   (s^2-2R')
   \left\{ \half (a')^2 - 2s[s\mp\sqrt{s^2-2R'}]a - Ns^2[s\mp\sqrt{s^2-2R'}]^2 
   \right\} \ .
\label{gauss-ode-a}
\end{multline}
Firstly the square-root appearing above is well defined as
\begin{equation}
 s^2-2R' = (s-2\tilde{R})^2 \ .
\label{gauss-sqr}
\end{equation}
One now faces a decision regarding the choice of the branch to be taken 
for the square root of the right-hand side of (\ref{gauss-ode-a}), and if 
we take the negative branch then the correct choice is given in 
(\ref{gauss-ode}). This choice is the only acceptable one given the 
boundary conditions (\ref{gauss-bc}) which impose that all quantities
$ a, R, \tilde{R} $ and derivatives vanish exponentially fast as 
$ s \to \infty $. If one takes the upper, negative sign then $ s-h $ also
vanishes in the same way and is consistent with the ODE (\ref{gauss-ode-a})
but if the other choice is made then $ s+h = {\rm O}(s) $ in this limit and
is incompatible with the ODE, both in the leading order $ s $-dependence and
the sign of the coefficient.
$ \square $

\subsection{Special Cases}

Here we calculate (\ref{E-gue}) in the cases $ N = 1 $ and $ N = 2 $ 
directly from the expansion (\ref{prob-0}, \ref{n-particle}, \ref{kernel}), 
which truncates after term $ n = N $. The formulas (\ref{gauss-sde:a}) and
(\ref{gauss-sqr}) then allow $ R(s) $ and $ \tilde{R}(s) $ to be computed
in terms of the error function.
\begin{proposition}
Consider the probability (\ref{E-gue}) and the associated functions $ R(s) $
as specified by (\ref{gauss-sde:a}) and $ \tilde{R}(s) $ as specified by
(\ref{tilde-R}). For $ N = 1 $, with the the standard definition of the error
function $ \erf(s) $, we have 
\begin{equation}
 \begin{split}
  E_2(0;I)
  & = \erf(s) \ , \\
  R(s)
  & = {e^{-s^2} \over \pi^{1/2}\erf(s)} \ , \\
  \tilde{R}(s)
  & = -{e^{-s^2} \over \pi^{1/2}\erf(s)} \ , \\
\end{split}
\label{gauss-N1}
\end{equation}
while for $ N = 2 $
\begin{equation}
 \begin{split}
  E_2(0;I)
  & = \erf(s)\left[ \erf(s) - 2\pi^{-1/2}se^{-s^2} \right] \ , \\
  R(s)
  & = {e^{-s^2} \over \pi^{1/2}\erf(s)}
      + {s^2e^{-s^2} \over \half\pi^{1/2}\erf(s)-se^{-s^2}} \ , \\
  \tilde{R}(s)
  & = {e^{-s^2} \over \pi^{1/2}\erf(s)}
      - {s^2e^{-s^2} \over \half\pi^{1/2}\erf(s)-se^{-s^2}} \ . \\
\end{split}
\label{gauss-N2}
\end{equation}
\end{proposition}
Using the kernel above, and (\ref{phi-psi}) and (\ref{hermite-poly}), 
the results follow from direct evaluation of the integrals. $ \square $

We can check (by hand and using computer algebra) that these expressions 
satisfy the appropriate equations (\ref{gauss-ode}, \ref{gauss-ode-b}) in 
Proposition 4, and furthermore satisfy the appropriate boundary conditions 
$ s \to \infty $, equation (\ref{gauss-bc}).

For small $ s $ we find from (\ref{gauss-N1}, \ref{gauss-N2}) that
\begin{equation}
 \begin{split}
  R^{N=1}
  & \sim  \frac{1}{2s}-\frac{1}{3}s+\frac{4}{45}s^3-\frac{8}{945}s^5
         -\frac{16}{14175}s^7+\frac{32}{93555}s^9
         +\frac{1472}{638512875}s^{11} \ , \\
  R^{N=2}
  & \sim  \frac{2}{s}-\frac{14}{15}s+\frac{248}{1575}s^3-\frac{128}{23625}s^5
         -\frac{51104}{27286875}s^7+\frac{1356032}{5320940625}s^9
         +\frac{6898816}{558698765625}s^{11} \ . \\
\end{split}
\label{gauss-N12-asymp}
\end{equation}
We can also determine the small-$s$ expansion for general $ N $. 
The form of this expansion can be understood from the small $ s $ behaviour
of (\ref{E-gue}). Now
\begin{align}
     E_{2}(0;(-\infty,-s)\cup(s,\infty);e^{-\lambda^2};N) 
   & = {1 \over C} \prod^{N}_{l=1} \int^{s}_{-s} d\lambda_{l}
       e^{-\lambda^2_{l}} \prod_{j < k} |\lambda_{k}-\lambda_{j}|^2
   \nonumber \\*
   & = {s^{N^2} \over C} \prod^{N}_{l=1} \int^{1}_{-1} d\lambda_{l}
       e^{-s^2\lambda^2_{l}} \prod_{j < k} |\lambda_{k}-\lambda_{j}|^2
   \nonumber \\*
   & \sim a_{0}s^{N^2} + a_{2}s^{N^2+2} + \ldots \ ,
\label{small-s-why}
\end{align}
where the final equality follows by expanding the exponential. According to
(\ref{gauss-sde:a}) and (\ref{gauss-sqr}) this implies 
\begin{equation}
   R(s) \sim {b_{-1} \over s} + b_{1}s + b_{3}s^3 + \ldots \ .
\label{small-s-exp}
\end{equation}
Making an ansatz of this form we find that all the coefficients are uniquely
determined by (\ref{gauss-ode}). Use of computer algebra gives the expansion
in the following result.
\begin{proposition}
For fixed $ N $, the asymptotic expansion of $ R(s) $ for small $ s $ is given 
by
\begin{align}
  R(s) & \sim 
   \frac{N^{2}}{2s}
  -\frac{N(2N^{2}-1)}{4N^{2}-1}s
  +\frac{2N^{2}(4N^{4}-9N^{2}+3)}
           {(4{N}^{2}-1)^{2}(4{N}^{2}-9)}s^{3} \nonumber \\*
  & \qquad
  +\frac{8N^{3}(4{N}^{4}-13{N}^{2}+6)}
           { (4{N}^{2}-1)^{3} (4{N}^{2}-9)
                    (4{N}^{2}-25)}s^{5} \nonumber \\*
  & \qquad\qquad
  +\frac{8N^{2}(128{N}^{10}-1312{N}^{8}+3304{N}^{6}
                    -3430{N}^{4}+1355{N}^{2}-315)}
           {(4{N}^{2}-1)^{4}(4{N}^{2}-9)^{2}
            (4{N}^{2}-25)(4{N}^{2}-49)}s^{7} \ .
\label{gauss-small-s}
\end{align}
\end{proposition}

It is of interest to note that in fact the only Laurent series about
$ s = 0 $ which satisfies (\ref{gauss-ode}) has the structure 
(\ref{gauss-small-s}). To see this it is simplest to consider the equation
(\ref{gauss-ode-b}). Making the Ansatz
\begin{equation}
  s\tilde{R} = \sum_{n=k}^{\infty} r_n s^n \ ,
\label{gauss-exp}
\end{equation}
we find that with a lower exponent $ k $ and $ k \geq 1 $ the only
solution for the coefficients is the null solution. However if $ k \leq -1 $
then one finds $ r_{k} = \ldots = r_{-1} = 0 $, and furthermore the relations
obtained by equating the coefficients of $ s^{-2}, s^{-1}, \ldots $ are the 
same as for the case of $ k = 0 $. Explicitly equating coefficients of 
$ s^{-2} $ in the ODE yields
\begin{equation}
  r^2_{1} + 8Nr^2_{0} + 16r^3_{0} = 0 \ .
\end{equation}
Now $ r_{0} \neq 0 $, therefore the odd-index terms must vanish because
$ s\tilde{R}(s) $ is even, as is seen from the ODE (\ref{gauss-ode-b}).
Thus $ r_{1} = 0 $ and an unique solution is found for $ r_{0} = -N/2 $. 
The coefficient of $ s^{-1} $ automatically vanishes while equating coefficients
of $ s^{0} $ gives
\begin{equation}
  r_{2} = { N^2 \over 4N^2\!-\!1 } \quad\text{or}\quad
           -N^2 \ .
\end{equation}
Choosing the former root, we find by equating coefficients of higher powers of
$ s $ that each of $ r_{3}, r_{4}, \ldots $ is now uniquely determined.

\subsection{Edge Scaling}
To leading order the support of the GUE is within the interval 
$ (-\sqrt{2N},\sqrt{2N}) $. The statistical properties of the eigenvalues 
in the vicinity of the edge ($ \lambda \sim \sqrt{2N} $ say) can be studied
by introducing the scale
\begin{equation}
   \lambda \mapsto \sqrt{2N} + {\lambda \over \sqrt{2}N^{1/6}} \ ,
\label{gauss-edge}
\end{equation}
which for $ N \to \infty $ makes the separation between the largest and the
second largest eigenvalue of order unity. In (\ref{E-gue}) the probability
that the region $ -\infty $ to the vicinity of the lowest eigenvalue, and
the highest eigenvalue to $ \infty $, is free of eigenvalues can be studied
in this limit by writing
\begin{equation}
  s = \sqrt{2N} + {t \over \sqrt{2}N^{1/6}} \ .
\label{gauss-edge-s}
\end{equation}
Because the two intervals in (\ref{E-gue}) become independent in this limit
we would expect
\begin{multline}
   \lim_{N \to \infty} 
   E_{2}(0;(-\infty,-(\sqrt{2N}+{t \over \sqrt{2}N^{1/6}})) \cup
           (\sqrt{2N}+{t \over \sqrt{2}N^{1/6}},\infty); e^{-\lambda^2};N) = \\
   \left( E^{\rm{soft}}_{2}(0;(t,\infty)) \right)^2 \ ,
\label{gauss-Elimit}
\end{multline}
where $ E^{\rm{soft}}_{2}(0;(t,\infty)) $ refers to the probability that the 
edge $ (t,\infty) $ is free of eigenvalue in the infinite GUE scaled
according to (\ref{gauss-edge}). Now it follows from (\ref{gauss-sde:a}) that
\begin{equation}
   E_{2}(0;(-\infty,-s)\cup(s,\infty);e^{-\lambda^2};N)
   = \exp\left( -2\int^{\infty}_{s}du R(u) \right) \ .
\label{gauss-Rlimit}
\end{equation}
On the other hand it is known that \cite{TW-94a}
\begin{equation}
   E^{\rm{soft}}_{2}(0;(t,\infty)) = 
   \exp\left( -\int^{\infty}_{t}du R^{\rm{soft}}(u) \right) \ ,
\label{gauss-softR}
\end{equation}
where $ R^{\rm{soft}}(t) $ satisfies 
\begin{equation}
  \left( \ddot{R}^{\rm{soft}} \right)^2
  + 4 \dot{R}^{\rm{soft}} \left[ \left( \dot{R}^{\rm{soft}} \right)^2
                           -t\dot{R}^{\rm{soft}} + R^{\rm{soft}} \right] = 0 \ ,
\label{gauss-JMO}
\end{equation}
which is the Jimbo-Miwa-Okamoto form of a particular \mbox{Painlev\'e-II}
equation, subject to the boundary conditions
\begin{equation}
   R^{\rm{soft}}(t) \underset{t \to \infty}{\sim}
  -t \left[ \Ai(t) \right]^2 + \left[ \Ai'(t) \right]^2 \ .
\label{gauss-JMO-bc}
\end{equation}
Thus for (\ref{gauss-Elimit}) to be valid we must have 
\begin{equation}
   \lim_{N \to \infty} {1 \over \sqrt{2}N^{1/6}}
    R(\sqrt{2N} + {t \over \sqrt{2}N^{1/6}}) = R^{\rm{soft}}(t)
\label{gauss-Rcompare}
\end{equation}
The equation (\ref{gauss-JMO}) can be verified by putting 
\begin{equation}
   r(t) = {1 \over \sqrt{2}N^{1/6}} R(\sqrt{2N} + {t \over \sqrt{2}N^{1/6}}) \ ,
\label{gauss-scale1}
\end{equation}
and showing that for $ N \to \infty $ $ r(t) $ satisfies the differential
equation (\ref{gauss-JMO}) and the boundary condition (\ref{gauss-JMO-bc}).
Regarding the latter point, note that (\ref{gauss-asympt}) implies 
\begin{equation}
   r(t) \underset{t \to \infty}{\sim} {1 \over \sqrt{2}N^{1/6}}
   K_N(\sqrt{2N} + {t \over \sqrt{2}N^{1/6}},
       \sqrt{2N} + {t \over \sqrt{2}N^{1/6}}) \ ,
\label{gauss-soft}
\end{equation}
where $ K_{N}(x,y) $ is specified by substituting (\ref{hermite-poly})
and (\ref{phi-psi}) into (\ref{kernel}). But we have the known limiting
behaviour \cite{F-00}
\begin{multline}
   \lim_{N \to \infty} {1 \over \sqrt{2}N^{1/6}}
   K_{N}(\sqrt{2N} + {X \over \sqrt{2}N^{1/6}},
         \sqrt{2N} + {Y \over \sqrt{2}N^{1/6}}) = \\
   K^{\rm{soft}}(X,Y) = {\Ai(X)\Ai'(Y) - \Ai(Y)\Ai'(X) \over X-Y}
\label{kernel-scale}
\end{multline}
which immediately implies (\ref{gauss-JMO-bc}). Thus it remains to show that
for $ N \to \infty $ (\ref{gauss-scale1}) satisfies (\ref{gauss-JMO}).

In fact it is more convenient to work with the ODE (\ref{gauss-ode-b}), and
introduce the scaled quantity
\begin{equation}
  \tilde{r}(t) = {N^{1/6} \over \sqrt{2}} 
  \tilde{R}(\sqrt{2N} + {t \over \sqrt{2}N^{1/6}}) \ .
\label{gauss-scale2}
\end{equation}
(this choice of scaling is consistent with the relation (\ref{gauss-sde:h})
between $ R(s) $ and $ \tilde{R}(s) $). With this substitution one finds that
for $ N \to \infty $ (\ref{gauss-ode-b}) reduces to
\begin{equation}
  2\tilde{r}\ddot{\tilde{r}} -(\dot{\tilde{r}})^2 + 16\tilde{r}^3  
  -4t \tilde{r}^2 = 0 \ ,
\label{gauss-largeN-ode1}
\end{equation}
while the relation (\ref{gauss-sde:h}) reduces to
\begin{equation}
  \dot{r} = 2\tilde{r} \ .
\label{gauss-largeN-ode2}
\end{equation}
Substituting (\ref{gauss-largeN-ode2}) in (\ref{gauss-largeN-ode1}) gives
the third order ODE
\begin{equation}
  2\dot{r}\dddot{r} - (\ddot{r})^2 + 8(\dot{r})^3 - 4t\dot{r}^2 = 0 \ .
\label{gauss-corres}
\end{equation}
Differentiating (\ref{gauss-JMO}) and using the original equation once more, 
it is easy to reduce it to (\ref{gauss-corres}), thus establishing the
required result. Alternatively one prove the same correspondence by making the 
substitution
\begin{equation}
  \tilde{r} = -\half q^2 \ ,
\label{gauss-xfm}
\end{equation}
in the ODE (\ref{gauss-largeN-ode1}), and then show this leads to a 
\mbox{Painlev\'e-II} equation with $ \alpha = 0 $, namely
\begin{equation}
  \ddot{q} = 2q^3+tq \ .
\label{gauss-PII}
\end{equation}
This equation has been derived for the kernel (\ref{kernel-scale}) in
\cite{TW-94a}).

\vfill\eject
\setcounter{equation}{0}

\section{The Jacobi Ensemble}
In this part of our study we will treat the Jacobi ensemble, for general
$ \alpha, \beta $ in some parts, but mostly we will consider only
the symmetrical case $ \alpha = \beta $ with our objective to compute 
the quantities (\ref{E-jue}). For the Jacobi weight in (\ref{weights})
the orthonormal polynomials are given in terms of the standard Jacobi 
polynomials by
\begin{equation}
 p_N(x) = \left[ {2N\!+\!\alpha\!+\!\beta\!+\!1 \over
                  2^{\alpha\!+\!\beta\!+\!1}}
                 {N!\Gamma(N\!+\!\alpha\!+\!\beta\!+\!1) \over 
                    \Gamma(N\!+\!\alpha\!+\!1)\Gamma(N\!+\!\beta\!+\!1)}
          \right]^{1/2} P^{(\alpha,\beta)}_{N}(x) \ ,
\label{jacobi-poly}
\end{equation}
with the corresponding coefficient of $ x^N $ such that
\begin{equation}
  {a_{N-1} \over a_{N}} =
  2\left[ { N(N\!+\!\alpha)(N\!+\!\beta)(N\!+\!\alpha\!+\!\beta) \over 
            (2N\!+\!\alpha\!+\!\beta)^2(2N\!+\!\alpha\!+\!\beta\!+\!1)
            (2N\!+\!\alpha\!+\!\beta\!-\!1) } \right]^{1/2} \ .
\label{jacobi-coeff}
\end{equation}
Making use of the differentiation formula
\begin{multline}
  (2N\!+\!\alpha\!+\!\beta)(1\!-\!x^2){d \over dx}P^{(\alpha,\beta)}_{N}(x)
  = N[\alpha\!-\!\beta\!-\!(2N\!+\!\alpha\!+\!\beta)x]P^{(\alpha,\beta)}_{N}(x)
  \\
  + 2(N\!+\!\alpha)(N\!+\!\beta)P^{(\alpha,\beta)}_{N-1}(x) \ ,
\label{jacobi-diff}
\end{multline}
and the three term recurrence for the Jacobi polynomials gives that the coupled
first order equations hold with \cite{TW-94}
\begin{equation}
 \begin{split}
  m(x)
  & = 1\!-\!x^2 \ , \\
  A(x)
  & = {\beta^2\!-\!\alpha^2 \over 2(2N\!+\!\alpha\!+\!\beta)}
      - {2N\!+\!\alpha\!+\!\beta \over 2}x = \alpha_0+\alpha_1 x  \ , \\
  B(x)
  & = {2\sqrt{N(N\!+\!\alpha)(N\!+\!\beta)(N\!+\!\alpha\!+\!\beta)} 
        \over 2N\!+\!\alpha\!+\!\beta}
      \sqrt{ 2N\!+\!\alpha\!+\!\beta\!+\!1 \over 2N\!+\!\alpha\!+\!\beta\!-\!1 }
    = \beta_0 \ , \\
  C(x)
  & = {2\sqrt{N(N\!+\!\alpha)(N\!+\!\beta)(N\!+\!\alpha\!+\!\beta)} 
        \over 2N\!+\!\alpha\!+\!\beta}
      \sqrt{ 2N\!+\!\alpha\!+\!\beta\!-\!1 \over 2N\!+\!\alpha\!+\!\beta\!+\!1 }
    = \gamma_0 \ . \\
\end{split}
\label{jacobi-param}
\end{equation}

\subsection{Differential Equations for End Intervals}
We will begin by considering the first probability in (\ref{E-jue}),
generalised so that the weight function is the general (nonsymmetric) Jacobi
form in (\ref{weights}). We shall adopt the conventions 
$ q_3,p_3 = q_{+},p_{+} $, $ q_2,p_2 = q_{-},p_{-} $ and 
$ R_{+} = R(s,s), R_{-} = R(-s,-s), R_{0} = R(-s,s) $.
\begin{proposition}
The coupled differential equations for the finite $ N $ JUE on the interval 
$ (-1,-s) \cup (s,1) $ for general $ \alpha, \beta $ are
\begin{align}
  [\ln E_{2}]'
  & = R_{-}+R_{+} \ ,
  \label{Ejacobi-sde:a}\\
  R_{0} 
  & = { q_{+}p_{-}-q_{-}p_{+} \over 2s } \ ,
  \label{Ejacobi-sde:b}\\
  (1\!-\!s^2)R_{-}
  & = [\gamma_0-w(2\alpha_1\!+\!1)] q^2_{-}
     + [\beta_0+u(2\alpha_1\!-\!1)] p^2_{-}
     + [\alpha_0-\alpha_1 s+v] 2q_{-}p_{-}
  \notag\\
  & \qquad + 2s(1\!-\!s^2)R^2_{0}  \ ,
  \label{Ejacobi-sde:c}\\
  (1\!-\!s^2)R_{+}
  & = [\gamma_0-w(2\alpha_1\!+\!1)] q^2_{+}
     + [\beta_0+u(2\alpha_1\!-\!1)] p^2_{+}
     + [\alpha_0+\alpha_1 s+v] 2q_{+}p_{+}
  \notag\\
  & \qquad + 2s(1\!-\!s^2)R^2_{0}  \ ,
  \label{Ejacobi-sde:d}\\
  u'
  & = -q^2_{-}-q^2_{+} \ ,
  \label{Ejacobi-sde:e}\\
  v'
  & = -q_{-}p_{-}-q_{+}p_{+} \ ,
  \label{Ejacobi-sde:f}\\
  w'
  & = -p^2_{-}-p^2_{+} \ ,
  \label{Ejacobi-sde:g}\\
  (1\!-\!s^2)q'_{-}
  & = - [\alpha_0-\alpha_1 s+v] q_{-}
      - [\beta_0+u(2\alpha_1\!-\!1)] p_{-}
      - 2(1\!-\!s^2)q_{+}R_{0} \ ,
  \label{Ejacobi-sde:h}\\
  (1\!-\!s^2)p'_{-}
  & = - [-\gamma_0+w(2\alpha_1\!+\!1)] q_{-}
      - [-\alpha_0+\alpha_1 s-v] p_{-}
      - 2(1\!-\!s^2)p_{+}R_{0} \ ,
  \label{Ejacobi-sde:i}\\
  (1\!-\!s^2)q'_{+}
  & =   [\alpha_0+\alpha_1 s+v] q_{+}
      + [\beta_0+u(2\alpha_1\!-\!1)] p_{+}
      - 2(1\!-\!s^2)q_{-}R_{0} \ ,
  \label{Ejacobi-sde:j}\\
  (1\!-\!s^2)p'_{+}
  & =   [-\gamma_0+w(2\alpha_1\!+\!1)] q_{+}
      + [-\alpha_0-\alpha_1 s-v] p_{+}
      - 2(1\!-\!s^2)p_{-}R_{0} \ ,
  \label{Ejacobi-sde:k}\\
  \left[(1\!-\!s^2)R_{-}\right]'
  & =  -2\alpha_1 q_{-}p_{-} - 2(1\!-\!s^2)R^2_{0} \ ,
  \label{Ejacobi-sde:l}\\
  \left[(1\!-\!s^2)R_{+}\right]'
  & =  +2\alpha_1 q_{+}p_{+} - 2(1\!-\!s^2)R^2_{0} \ .
  \label{Ejacobi-sde:m}
\end{align}
\end{proposition}
The first equation follows from (\ref{diff-E}), (\ref{prob-fd}) and the next
equality in (\ref{Ejacobi-sde:b}) follows from (\ref{fn-R}), 
(\ref{Ejacobi-sde:c}) and (\ref{Ejacobi-sde:d}) follow from (\ref{diag-R}),
while (\ref{Ejacobi-sde:l}) and (\ref{Ejacobi-sde:m}) follow from 
(\ref{diff-diag-R}). Furthermore (\ref{Ejacobi-sde:e})-(\ref{Ejacobi-sde:g})
follow from (\ref{diff-uvw}) and the first equality in (\ref{Ejacobi-sde:b})
and (\ref{Ejacobi-sde:h})-(\ref{Ejacobi-sde:k}) follow from 
(\ref{diff-diag-pq}) and (\ref{Ejacobi-sde:b}).
$ \square $

We note from (\ref{Ejacobi-sde:l}), (\ref{Ejacobi-sde:m}) and 
(\ref{Ejacobi-sde:f}) the integral
\begin{equation}
  (1\!-\!s^2)(R_{-}-R_{+}) = 2\alpha_1 v \ .
\label{Ejacobi-int}
\end{equation}
The boundary conditions satisfied by $ R(s,s) $ as $ s \to 1^{-} $ or
$ s \to -1^{+} $ can be expressed as
\begin{align}
  R(s,s) \sim
  &  {N!\Gamma(N\!+\!\alpha\!+\!\beta\!+\!1) \over 
        \Gamma(N\!+\!\alpha)\Gamma(N\!+\!\beta)}
         {(1\!-\!s)^{\alpha-1}(1\!+\!s)^{\beta-1} \over
           2^{\alpha+\beta}(2N\!+\!\alpha\!+\!\beta)}
    \nonumber \\*
  & \qquad \times
     \left[ \left({ \beta^2\!-\alpha^2 \over 
                    (2N\!+\!\alpha\!+\!\beta)(2N\!+\!\alpha\!+\!\beta\!+\!2) }
                   -s \right)
            P^{(\alpha,\beta)}_{N}(s)P^{(\alpha,\beta)}_{N-1}(s)
     \right. \nonumber \\*
  & \qquad\qquad
            -{ 2(N\!+\!1)(N\!+\!\alpha\!+\!\beta\!+\!1) \over 
               2N\!+\!\alpha\!+\!\beta\!+\!2 }
             P^{(\alpha,\beta)}_{N+1}(s)P^{(\alpha,\beta)}_{N-1}(s)
     \nonumber \\*
  & \qquad\qquad\qquad\left. 
            +{ 2N(N\!+\!\alpha\!+\!\beta) \over 
               2N\!+\!\alpha\!+\!\beta }
              \left[ P^{(\alpha,\beta)}_{N}(s) \right]^2
     \right] \ .
\label{Ejacobi-bc}
\end{align}

Our objective is to use equations such as (\ref{Ejacobi-int}) to reduce the
equations of Proposition 7 down to a single equation for $ R_{+}(s) $.
For this purpose we restrict attention to the case $ \alpha = \beta $,
then we see from (\ref{jacobi-param}) that $ \alpha_0 = 0 $, while the fact
that
\begin{equation}
  P^{(\alpha,\alpha)}_{N}(-x) = (-1)^{N}P^{(\alpha,\alpha)}_{N}(x) \ ,
\label{jacobi-parity}
\end{equation}
implies $ v = 0 $, $ q_{2} = (-1)^{N}q_{3} $, $ p_{2} = (-1)^{N-1}p_{3} $
and $ R_{+} = R_{-} $ ($ = R $ say), thus reducing the number of unknowns in
Proposition 7.
\begin{proposition}
The coupled differential equations for the finite $ N $ JUE on the interval 
$ (-1,-s) \cup (s,1) $ for $ \beta = \alpha $ are
\begin{align}
  [\ln E_{2}]'
  & = 2R \ ,
  \label{Ejacobi-eq-sde:a}\\
  R(-s,s)
  & = (-1)^{N-1}{ qp \over s } \equiv (-1)^{N-1}R_{0} \ ,
  \label{Ejacobi-eq-sde:b}\\
  (1\!-\!s^2)R
  & = [\gamma_0-w(2\alpha_1\!+\!1)] q^2
     + [\beta_0+u(2\alpha_1\!-\!1)] p^2
     + 2\alpha_1 sqp
     + 2s(1\!-\!s^2)R^2_{0}  \ ,
  \label{Ejacobi-eq-sde:c}\\
  u'
  & = -2q^2 \ ,
  \label{Ejacobi-eq-sde:d}\\
  w'
  & = -2p^2 \ ,
  \label{Ejacobi-eq-sde:e}\\
  (1\!-\!s^2)q'
  & = + \alpha_1 sq
      + [\beta_0+u(2\alpha_1\!-\!1)] p
      + {2(1\!-\!s^2) \over s}q^2p \ ,
  \label{Ejacobi-eq-sde:f}\\
  (1\!-\!s^2)p'
  & = - \alpha_1 sp
      + [-\gamma_0+w(2\alpha_1\!+\!1)] q
      - {2(1\!-\!s^2) \over s}qp^2 \ ,
  \label{Ejacobi-eq-sde:g}\\
  \left[(1\!-\!s^2)R\right]'
  & =  2\alpha_1 qp - {2(1\!-\!s^2) \over s^2}q^2p^2 \ ,
  \label{Ejacobi-eq-sde:h}
\end{align}
where we have redefined $ R_{0} $ from the previous usage 
($ R_{0} \mapsto (-1)^{N-1}R_{0} $) and made the notation 
$ q_{+} = q, p_{+} = p $.
\end{proposition}

We now indicate how to reduce such a system to a single third order 
differential equation for $ R = R(s) = R(s,s) $.
\begin{proposition}
The coupled set of ODEs given in Proposition 8 reduce to the third order 
ODE for $ \sigma(s) = (1\!-\!s^2)R(s) $,
\begin{align}
 & (1\!-\!s^2)^2\sigma''' - 2s(1\!-\!s^2)\sigma''
       - 2s^{-2}(1\!-\!s^2)\sigma' - (1\!-\!s^2)^2(\sigma'')^2/\sigma'
 \nonumber \\*
 & \qquad
   + \alpha_1 {s \over 2\sigma'}{\alpha_1 s + H \over H^2}
     \left[ (1\!-\!s^2)\sigma'' - 2s^{-1}\sigma' \right]^2
 \nonumber \\*
 & \qquad
   + 4\alpha_1 H\sigma + 2[\alpha_1 s + H](\alpha_1 s^{-1} - H\sigma')
     + 2{\sigma^2 \over s^2\sigma'}H[H - \alpha_1 s]
  = 0 \ ,
\label{Ejacobi-ode3}
\end{align}
where $ H \equiv \sqrt{\alpha^2_1 s^2 - 2(1\!-\!s^2)\sigma'} $.
\end{proposition}
Proof - Firstly we make some auxiliary definitions similar to the Gaussian case
\begin{equation}
 \begin{split}
  a & = qp \ , \\
  A & =   [\beta_0+u(2\alpha_1\!-\!1)] p^2 \ , \\
  B & = [-\gamma_0+w(2\alpha_1\!+\!1)] q^2 \ . \\
\end{split}
\label{Ejacobi-aux}
\end{equation}
Combining the relations (\ref{Ejacobi-eq-sde:f}) and (\ref{Ejacobi-eq-sde:g}) 
we find
\begin{equation}
   (1\!-\!s^2)a' = A+B \ ,
\label{Ejacobi-elim1}
\end{equation}
and we differentiate this once more, using the relations 
(\ref{Ejacobi-eq-sde:d}), (\ref{Ejacobi-eq-sde:e}) 
(\ref{Ejacobi-eq-sde:f}) and (\ref{Ejacobi-eq-sde:g}), to arrive at
\begin{align}
  (1\!-\!s^2)^2a'' = 
 & 2s(A+B) - 8\alpha_1(1\!-\!s^2)a^2 
 \nonumber \\*
 &  \qquad\qquad
   + [2\alpha_1 s+4s^{-1}(1\!-\!s^2)a](B-A) + 4AB/a \ .
\label{Ejacobi-elim2}
\end{align}
Now the idea is to express $ A $, $ B $ in terms of $ \sigma $ and
its derivatives, so we employ (\ref{Ejacobi-elim1}) above and the relation
\begin{equation}
   s\sigma' + \sigma - 4\alpha_1 sa = A-B \ ,
\label{Ejacobi-elim3}
\end{equation}
which follows from (\ref{Ejacobi-eq-sde:c}) and (\ref{Ejacobi-eq-sde:h}). 
In order to express $ a $ in terms of $ \sigma'(s) $ we have to solve the 
quadratic relation (\ref{Ejacobi-eq-sde:h}) for $ a $ and this is how the 
square-root variable $ H $ arises.  This quantity is well defined because
\begin{equation}
  \alpha^2_1 s^2 - 2(1\!-\!s^2)\sigma' = 
  \left[ 2{1\!-\!s^2 \over s}a - \alpha_1 s \right]^2 \ .
\label{Ejacobi-sqr}
\end{equation}
Having expressed $ A $, $ B $ and $ a $ in terms of $ \sigma $ and it 
derivatives it is then a matter of substituting these into (\ref{Ejacobi-elim2})
and after considerable simplification we find the final result 
(\ref{Ejacobi-ode3}).
$ \square $

\subsection{Special Cases for End Intervals}
In this part we present the calculations for the first two finite-$ N $ cases,
that is $ N = 1 $ and $ N = 2 $, by direct means using the probability 
$ E_2(0;I) $.
\begin{proposition}
The probability that there are no eigenvalues in the interval
$ I = (-1,-s)\cup(s,1) $ of the JUE with $ N = 1 $, and general 
$ \alpha \neq \beta $ is
\begin{equation}
  E_2(0;I) =  I_{(1+s)/2}(\alpha\!+\!1,\beta\!+\!1)
            - I_{(1-s)/2}(\alpha\!+\!1,\beta\!+\!1)
\label{Ejacobi-N1}
\end{equation}
and for $ N = 2 $, is
\begin{align}
  E_2(0;I) =  
  & \phantom{-}(\alpha\!+\!\beta\!+\!3)
    \left[ I_{(1+s)/2}(\alpha\!+\!1,\beta\!+\!2)
                             - I_{(1-s)/2}(\alpha\!+\!1,\beta\!+\!2) \right]
  \nonumber \\*
  & \phantom{-(\alpha\!+\!\beta\!+\!3)} \times
      \left[ I_{(1+s)/2}(\alpha\!+\!2,\beta\!+\!1)
                             - I_{(1-s)/2}(\alpha\!+\!2,\beta\!+\!1) \right]
  \nonumber \\*
  & -(\alpha\!+\!\beta\!+\!2)
    \left[ I_{(1+s)/2}(\alpha\!+\!2,\beta\!+\!2)
                             - I_{(1-s)/2}(\alpha\!+\!2,\beta\!+\!2) \right]
  \nonumber \\*
  & \phantom{-(\alpha\!+\!\beta\!+\!2)} \times
    \left[ I_{(1+s)/2}(\alpha\!+\!1,\beta\!+\!1)
                             - I_{(1-s)/2}(\alpha\!+\!1,\beta\!+\!1) \right]
\label{Ejacobi-N2}
\end{align}
where the normalised incomplete beta functions are defined by
\begin{equation}
  I_{x}(a,b) = {B_{x}(a,b) \over B(a,b)}
\label{incompl-beta}
\end{equation}
in terms of the incomplete and complete beta functions, 
$ B_{x}(a,b) $ and $ B(a,b) $ respectively.
\end{proposition}
Proof - as in the case of Proposition 5, this follows from the expansion
(\ref{prob-0}) with (\ref{n-particle}) which truncates after the term $ n = N $
used in conjunction with (\ref{kernel}), (\ref{phi-psi}), 
(\ref{jacobi-poly}) and (\ref{jacobi-coeff}). Extensive use of the identities
for the incomplete beta function
\begin{equation}
  \begin{split}
     I_{x}(a,b) & = 1 - I_{1-x}(b,a) \ , \notag\\
     (a\!+\!b)I_{x}(a,b) & = aI_{x}(a\!+\!1,b) + bI_{x}(a,b\!+\!1) \ , \notag\\
  \end{split}
\end{equation}
are made to reduce the number of their occurrences.
$ \square $

In the symmetric case $ \alpha = \beta $ considerable simplification ensues 
and we have the following results.
\begin{proposition}
In the case $ \beta = \alpha $, the probability (\ref{Ejacobi-N1}) and 
the associated quantities of Proposition 10 are given in terms of the Gauss 
hypergeometric function $ {}_{2}F_{1}(a,b;c;z) $ by
\begin{equation}
 \begin{split}
  E_2(0;I)
  & = {1 \over 4^{\alpha}B(\alpha\!+\!1,\alpha\!+\!1)}
       s\, {}_{2}F_{1}(-\alpha,\half;\thalf;s^2)  \ , \\
  \sigma(s)
  & = \half { (1\!-\!s^2)^{\alpha+1} \over 
       s\, {}_{2}F_{1}(-\alpha,\half;\thalf;s^2) } \ , \\
  H(s)
  & = (\alpha\!+\!1)s + { (1\!-\!s^2)^{\alpha+1} \over 
       s\, {}_{2}F_{1}(-\alpha,\half;\thalf;s^2) } \ , \\
\end{split}
\label{Ejacobi-eq-N1}
\end{equation}
and similarly for the probability (\ref{Ejacobi-N2}),
\begin{equation}
 \begin{split}
  E_2(0;I)
  & = {2\alpha\!+\!3 \over 4^{2\alpha+1}B^2(\alpha\!+\!1,\alpha\!+\!2)}
       \othird s^4\, {}_{2}F_{1}(-\alpha,\half;\thalf;s^2)\,
                     {}_{2}F_{1}(-\alpha,\thalf;\fhalf;s^2)  \ , \\
  \sigma(s)
  & = { (1\!-\!s^2)^{\alpha+1} \over 2s }
      \left\{
       {1 \over {}_{2}F_{1}(-\alpha,\half;\thalf;s^2) }
     + {3 \over {}_{2}F_{1}(-\alpha,\thalf;\fhalf;s^2) }
      \right\} \ , \\
  H(s)
  & = (\alpha\!+\!2)s + { (1\!-\!s^2)^{\alpha+1} \over s }
      \left\{
     - {1 \over {}_{2}F_{1}(-\alpha,\half;\thalf;s^2) }
     + {3 \over {}_{2}F_{1}(-\alpha,\thalf;\fhalf;s^2) }
      \right\} \ . \\
\end{split}
\label{Ejacobi-eq-N2}
\end{equation}
\end{proposition}
Proof - These follow from the reduction of the incomplete beta functions in the 
symmetric case to Hypergeometric functions, such as
\begin{equation}
   I_{(1+s)/2}(\alpha\!+\!1,\alpha\!+\!1) - 
   I_{(1-s)/2}(\alpha\!+\!1,\alpha\!+\!1) = 
   {1 \over 4^{\alpha}B(\alpha\!+\!1,\alpha\!+\!1)}
    s\, {}_{2}F_{1}(-\alpha,\half;\thalf;s^2) \ ,
\end{equation}
the differentiation formulae for these Hypergeometric functions, like
\begin{equation}
   {d \over ds}[ s\,{}_{2}F_{1}(-\alpha,\half;\thalf;s^2) ] =
   (1\!-\!s^2)^{\alpha}
\end{equation}
and the use of their contiguous relations
\begin{equation}
      (2\alpha\!+\!3)\othird s^3 {}_{2}F_{1}(-\alpha,\thalf;\fhalf;s^2)
    - s\, {}_{2}F_{1}(-\alpha,\half;\thalf;s^2)
    + s\, {}_{2}F_{1}(-\alpha\!-\!1,\thalf;\thalf;s^2) = 0 \ .
\label{Econtig-reln}
\end{equation}
$ \square $

One can show that these two specific cases are solutions to our third order
differential equation (\ref{Ejacobi-ode3}), after noting the contiguous 
relation above (\ref{Econtig-reln}) linking the two Hypergeometric functions 
in the case of $ N = 2 $.

Another special case for which the probability can be computed independently 
of the ODE (\ref{Ejacobi-ode3}) is that with $ \alpha = \beta = 0 $ and 
general $ N $. We then have
\begin{equation}
  E_{2}(0;(-1,-s)\cup(s,1);\chi_{[0,1]};N) = 
  {1 \over C} \int^{s}_{-s}d\lambda_{1} \cdots \int^{s}_{-s}d\lambda_{N}
  \prod_{1 \leq j < k \leq N}|\lambda_{j}-\lambda_{k}|^{2} \ ,
\label{Ejacobi-zero-int}
\end{equation}
where $ C $ is such that $ E_{2} = 1 $ for $ s = 1 $, which can be evaluated
by a change of variable.
\begin{proposition}
The probability (\ref{Ejacobi-zero-int}) and the associated quantities of
Proposition 10 have the evaluation
\begin{equation}
 \begin{split}
  E_2(0;I)
  & = s^{N^2} \ , \\
  \sigma(s)
  & = N^2{ (1\!-\!s^2) \over 2s } \ , \\
  H(s)
  & = { N \over s } \ . \\
\end{split}
\label{Ejacobi-zero}
\end{equation}
\end{proposition}
$ \square $

This exact form of $ \sigma(s) $ can be verified to satisfy 
(\ref{Ejacobi-ode3}) with $ \alpha = \beta = 0 $. Furthermore we can take the
limit $ \alpha \to 0 $ in (\ref{Ejacobi-eq-N1}) and (\ref{Ejacobi-eq-N2})
and reclaim (\ref{Ejacobi-zero}) in the cases $ N = 1 $ and $ N = 2 $
respectively.

\subsection{Jacobi to Hermite Limit for the End Intervals}
From the definition
\begin{multline}
   E_{2}(0;(-1,-s)\cup(s,1);(1\!-\!\lambda^2)^{\alpha};N) = \\
     {1 \over C}\int^{s}_{-s}d\lambda_{1} \cdots \int^{s}_{-s}d\lambda_{N}
     \prod^{N}_{l=1}(1\!-\!\lambda^2_{l})^{\alpha}
     \prod_{1 \leq j < k \leq N}|\lambda_{k}-\lambda_{j}|^2 \ ,
\label{Ejacobi-prob-int}
\end{multline}
where again the constant $ C $ ensures the normalisation. Replacing $ s $ 
by $ t/\sqrt{\alpha} $, changing variables 
$ \lambda_{l} \mapsto \lambda_{l}/\sqrt{\alpha} $ and taking
$ \alpha \to \infty $ shows
\begin{multline}
  E_{2}(0;(-1,-t/\sqrt{\alpha})\cup
          (t/\sqrt{\alpha},1);(1\!-\!\lambda^2)^{\alpha};N) 
  \underset{\alpha \to \infty}{\sim} \\
  E_{2}(0;(-\infty,-t)\cup(t,\infty);e^{-\lambda^2};N) \ .
\label{Ejacobi-j2g-scale}
\end{multline}
Recalling (\ref{gauss-sde:a}) and (\ref{Ejacobi-eq-sde:a}) this is 
equivalent to the statement that 
\begin{equation}
  \lim_{\alpha \to \infty}{1 \over \sqrt{\alpha}}R({t \over \sqrt{\alpha}})
  = R^{\rm{GUE}}(t) \ ,
\label{Ejacobi-j2g-R}
\end{equation}
where $ R(s) $ on the left-hand side is as in Propositions 7 and 8 while
$ R^{\rm{GUE}}(t) $ is the $ R $ defined in Proposition 3. Thus it must be that
with
\begin{equation}
  {1 \over \sqrt{\alpha}}R({t \over \sqrt{\alpha}}) = r(t) \ ,
\label{j2g-xfm}
\end{equation}
in (\ref{Ejacobi-ode3}), taking the limit $ \alpha \to \infty $ must take
the third order equation into an equation equivalent to the second order 
equation (\ref{gauss-ode}).

To verify this, we first note that the substitution (\ref{j2g-xfm}), to
leading order in $ \alpha^2 $ reads
\begin{align}
 & r''' + 2 - {(r'')^2 \over r'} - {2\over t^{2}}r'
   + {2 \over t^{2}}{r^2 \over r'} h[t+h]
 \nonumber \\*
 & \qquad
    + 2r' h[t-h] - \left(4r+{2 \over t}\right)h
    + {(tr''-2r')^2 \over 2tr'h^2}[t-h] = 0 \ ,
\label{j2g-ode3}
\end{align}
and that differentiation of the second order Hermite ODE, (\ref{gauss-ode}),
gives
\begin{align}
 tr''' + 3r'' = 
 & 4t + 4h^2(tr'+r+2Nt) - 2h 
 \nonumber \\*
 & - 8t(r+Nt)h - (t-r'')(2t^2-tr''-2r')/h^2 \ ,
\label{gauss-ode3}
\end{align}
with $ h = \sqrt{t^2-2r'} $, as before.
By effecting a suitable subtraction of these two equation in order to eliminate
the term in $ r''' $, and using the second order ODE once more, one can
show this difference is identically zero.

Employing this limit one can show that the special cases of $ N = 1 $ and 
$ N = 2 $ for the Jacobi weight (\ref{Ejacobi-eq-N1}) and (\ref{Ejacobi-eq-N2})
lead exactly to those for the Hermite weights (\ref{gauss-N1}) and 
(\ref{gauss-N2}), respectively.
 
\subsection{Differential Equations for an Interior Interval}
We will now consider our last case, the second probability in (\ref{E-jue}),
in a parallel manner to that of the previous case. Treating first the 
nonsymmetrical form $ \alpha \neq \beta $, we make the new conventions 
$ q_2,p_2 = q_{+},p_{+} $, $ q_1,p_1 = q_{-},p_{-} $ and 
$ R_{+} = R(s,s), R_{-} = R(-s,-s), R_{0} = R(-s,s) $.
\begin{proposition}
The coupled differential equations for the finite $ N $ JUE on the interval 
$ (-s,s) $ for general $ \alpha, \beta $ are
\begin{align}
  [\ln E_{2}]'
  & = -R_{-}-R_{+} \ ,
  \label{Ijacobi-sde:a}\\
  R_{0} 
  & = { q_{+}p_{-}-q_{-}p_{+} \over 2s } \ ,
  \label{Ijacobi-sde:b}\\
  (1\!-\!s^2)R_{-}
  & = [\gamma_0-w(2\alpha_1\!+\!1)] q^2_{-}
     + [\beta_0+u(2\alpha_1\!-\!1)] p^2_{-}
     + [\alpha_0-\alpha_1 s+v] 2q_{-}p_{-}
  \notag\\
  & \qquad - 2s(1\!-\!s^2)R^2_{0}  \ ,
  \label{Ijacobi-sde:c}\\
  (1\!-\!s^2)R_{+}
  & = [\gamma_0-w(2\alpha_1\!+\!1)] q^2_{+}
     + [\beta_0+u(2\alpha_1\!-\!1)] p^2_{+}
     + [\alpha_0+\alpha_1 s+v] 2q_{+}p_{+}
  \notag\\
  & \qquad - 2s(1\!-\!s^2)R^2_{0}  \ ,
  \label{Ijacobi-sde:d}\\
  u'
  & = +q^2_{-}+q^2_{+} \ ,
  \label{Ijacobi-sde:e}\\
  v'
  & = +q_{-}p_{-}+q_{+}p_{+} \ ,
  \label{Ijacobi-sde:f}\\
  w'
  & = +p^2_{-}+p^2_{+} \ ,
  \label{Ijacobi-sde:g}\\
  (1\!-\!s^2)q'_{-}
  & = - [\alpha_0-\alpha_1 s+v] q_{-}
      - [\beta_0+u(2\alpha_1\!-\!1)] p_{-}
      + 2(1\!-\!s^2)q_{+}R_{0} \ ,
  \label{Ijacobi-sde:h}\\
  (1\!-\!s^2)p'_{-}
  & = - [-\gamma_0+w(2\alpha_1\!+\!1)] q_{-}
      - [-\alpha_0+\alpha_1 s-v] p_{-}
      + 2(1\!-\!s^2)p_{+}R_{0} \ ,
  \label{Ijacobi-sde:i}\\
  (1\!-\!s^2)q'_{+}
  & =   [\alpha_0+\alpha_1 s+v] q_{+}
      + [\beta_0+u(2\alpha_1\!-\!1)] p_{+}
      + 2(1\!-\!s^2)q_{-}R_{0} \ ,
  \label{Ijacobi-sde:j}\\
  (1\!-\!s^2)p'_{+}
  & =   [-\gamma_0+w(2\alpha_1\!+\!1)] q_{+}
      + [-\alpha_0-\alpha_1 s-v] p_{+}
      + 2(1\!-\!s^2)p_{-}R_{0} \ ,
  \label{Ijacobi-sde:k}\\
  \left[(1\!-\!s^2)R_{-}\right]'
  & =  -2\alpha_1 q_{-}p_{-} + 2(1\!-\!s^2)R^2_{0} \ ,
  \label{Ijacobi-sde:l}\\
  \left[(1\!-\!s^2)R_{+}\right]'
  & =  +2\alpha_1 q_{+}p_{+} + 2(1\!-\!s^2)R^2_{0} \ .
  \label{Ijacobi-sde:m}
\end{align}
\end{proposition}
Proof - these follow in an entirely parallel manner as for the derivation of 
the previous set (\ref{Ejacobi-sde:a})-(\ref{Ejacobi-sde:m}).
$ \square $

The only, apparently minor, difference between the interior interval set of 
equations, (\ref{Ijacobi-sde:a})-(\ref{Ijacobi-sde:m}), and the endpoint 
interval set, (\ref{Ejacobi-sde:a})-(\ref{Ejacobi-sde:m}), is a change in sign 
of a number of terms in the expressions for derivatives. However, in reality,
the two cases are quite distinct.

Again we note that (\ref{Ijacobi-sde:l}), (\ref{Ijacobi-sde:m}) and 
(\ref{Ijacobi-sde:f}) imply the integral
\begin{equation}
  (1\!-\!s^2)(R_{+}-R_{-}) = 2\alpha_1 v \ .
\label{Ijacobi-int}
\end{equation}
The boundary conditions satisfied by $ R(s,s) $ now apply as $ s \to 0 $ 
and the limiting value takes the same form as (\ref{Ejacobi-bc}).

We continue by considering the symmetrical case $ \alpha = \beta $, and find
again that $ \alpha_0 = 0 $, and the parity relation (\ref{jacobi-parity}) 
implies $ v = 0 $ and $ R_{+} = R_{-} $, which is denoted by $ R $.
\begin{proposition}
The coupled differential equations for the finite $ N $ JUE on the interval 
$ (-s,s) $ for $ \beta = \alpha $ are
\begin{align}
  [\ln E_{2}]'
  & = -2R \ ,
  \label{Ijacobi-eq-sde:a}\\
  R(-s,s)
  & = (-1)^{N-1}{ qp \over s } = (-1)^{N-1}R_{0} \ ,
  \label{Ijacobi-eq-sde:b}\\
  (1\!-\!s^2)R
  & = [\gamma_0-w(2\alpha_1\!+\!1)] q^2
     + [\beta_0+u(2\alpha_1\!-\!1)] p^2
     + 2\alpha_1 sqp
     - 2s(1\!-\!s^2)R^2_{0}  \ ,
  \label{Ijacobi-eq-sde:c}\\
  u'
  & = +2q^2 \ ,
  \label{Ijacobi-eq-sde:d}\\
  w'
  & = +2p^2 \ ,
  \label{Ijacobi-eq-sde:e}\\
  (1\!-\!s^2)q'
  & = + \alpha_1 sq
      + [\beta_0+u(2\alpha_1\!-\!1)] p
      - {2(1\!-\!s^2) \over s}q^2p \ ,
  \label{Ijacobi-eq-sde:f}\\
  (1\!-\!s^2)p'
  & = - \alpha_1 sp
      + [-\gamma_0+w(2\alpha_1\!+\!1)] q
      + {2(1\!-\!s^2) \over s}qp^2 \ ,
  \label{Ijacobi-eq-sde:g}\\
  \left[(1\!-\!s^2)R\right]'
  & =  2\alpha_1 qp + {2(1\!-\!s^2) \over s^2}q^2p^2 \ ,
  \label{Ijacobi-eq-sde:h}
\end{align}
where we employ the notation $ q_{2} = q, p_{2} = p $. 
\end{proposition}

Again such a system can be reduced to a single third order differential 
equation for $ R(s) $.
\begin{proposition}
The coupled set of ODEs given in Proposition 13 are equivalent to the third 
order ODE for $ \sigma(s) = (1\!-\!s^2)R(s) $,
\begin{align}
 & (1\!-\!s^2)^2\sigma''' - 2s(1\!-\!s^2)\sigma''
       - 2s^{-2}(1\!-\!s^2)\sigma' - (1\!-\!s^2)^2(\sigma'')^2/\sigma'
 \nonumber \\*
 & \qquad
   + \alpha_1 {s \over 2\sigma'}{\alpha_1 s + G \over G^2}
     \left[ (1\!-\!s^2)\sigma'' - 2s^{-1}\sigma' \right]^2
 \nonumber \\*
 & \qquad
   + 4\alpha_1 G\sigma - 2[\alpha_1 s + G](\alpha_1 s^{-1} + G\sigma')
     + 2{\sigma^2 \over s^2\sigma'}G[G - \alpha_1 s]
  = 0 \ ,
\label{Ijacobi-ode3}
\end{align}
where $ G \equiv \sqrt{\alpha^2_1 s^2 + 2(1\!-\!s^2)\sigma'} $.
\end{proposition}
Proof - this proceeds in an identical manner to the proof of Proposition 9
but with a number of minor alterations. As in the previous reduction we have 
to express the following equation
\begin{align}
  (1\!-\!s^2)^2a'' = 
 & 2s(A+B) + 8\alpha_1(1\!-\!s^2)a^2 
 \nonumber \\*
 & \qquad\qquad
   + [2\alpha_1 s-4s^{-1}(1\!-\!s^2)a](B-A) + 4AB/a \ ,
\label{Ijacobi-elim}
\end{align}
in terms of $ \sigma $ and its derivatives alone. We have an analogous 
expression for $ \sigma' $ which is quadratic in $ a $ and its solution 
contains the square-root variable $ G $ defined above. This is well defined 
due to the relation
\begin{equation}
  \alpha^2_1 s^2 + 2(1\!-\!s^2)\sigma' = 
  \left[ 2{1\!-\!s^2 \over s}a + \alpha_1 s \right]^2 \ .
\label{Ijacobi-sqr}
\end{equation}
The final result is the above third order ODE (\ref{Ijacobi-ode3}).
$ \square $

\subsection{Special Cases for the Interior Interval}
In this part we present the analogous results for the first two finite-$ N $ 
cases $ N = 1 $ and $ N = 2 $ by direct calculation of the probability 
$ E_2(0;I) $.
\begin{proposition}
The probability that no eigenvalues are found in the interval $ (-s,s) $
for the JUE with $ N = 1 $, and general $ \alpha \neq \beta $ is
\begin{equation}
  E_2(0;I) = 1 - I_{(1+s)/2}(\alpha\!+\!1,\beta\!+\!1)
               + I_{(1-s)/2}(\alpha\!+\!1,\beta\!+\!1)
\label{Ijacobi-N1}
\end{equation}
and for $ N = 2 $, is
\begin{align}
  E_2(0;I) =  
  & \phantom{-}(\alpha\!+\!\beta\!+\!3)
    \left[ 1 + I_{(1-s)/2}(\alpha\!+\!1,\beta\!+\!2)
                             - I_{(1+s)/2}(\alpha\!+\!1,\beta\!+\!2) \right]
  \nonumber \\*
  & \phantom{-(\alpha\!+\!\beta\!+\!3)]} \times
    \left[ 1 + I_{(1-s)/2}(\alpha\!+\!2,\beta\!+\!1)
                             - I_{(1+s)/2}(\alpha\!+\!2,\beta\!+\!1) \right]
  \nonumber \\*
  & -(\alpha\!+\!\beta\!+\!2)
    \left[ 1 + I_{(1-s)/2}(\alpha\!+\!2,\beta\!+\!2)
                             - I_{(1+s)/2}(\alpha\!+\!2,\beta\!+\!2) \right]
  \nonumber \\*
  & \phantom{-(\alpha\!+\!\beta\!+\!2)} \times
    \left[ 1 + I_{(1-s)/2}(\alpha\!+\!1,\beta\!+\!1)
                             - I_{(1+s)/2}(\alpha\!+\!1,\beta\!+\!1) \right]
\label{Ijacobi-N2}
\end{align}
in terms of the normalised incomplete beta functions, $ I_{x}(a,b) $.
\end{proposition}
Proof - these follow from direct evaluations of the terminating series for the
probability $ E_{2}(0;I) $ given in (\ref{prob-0}) as indicated in the proof
of Proposition 10.
$ \square $

When equality $ \alpha = \beta $ holds then the simpler results follow - 
\begin{proposition}
For equal parameters $ \beta = \alpha $ the probability (\ref{Ijacobi-N1}) 
and associated functions arising from Proposition 16 are given by
\begin{equation}
 \begin{split}
  E_2(0;I)
  & = {s(1\!-\!s^2)^{\alpha+1} \over
       2^{2\alpha+1}(\alpha\!+\!1)B(\alpha\!+\!1,\alpha\!+\!1)}
       {}_{2}F_{1}(\alpha\!+\!\thalf,1;\alpha\!+\!2;1\!-\!s^2)
  \ , \\
  \sigma(s)
  & = { \alpha\!+\!1 \over 
       s\, {}_{2}F_{1}(\alpha\!+\!\thalf,1;\alpha\!+\!2;1\!-\!s^2) }
  \ , \\
  G(s)
  & = (\alpha\!+\!1)\left\{ s - { 2 \over 
       s\, {}_{2}F_{1}(\alpha\!+\!\thalf,1;\alpha\!+\!2;1\!-\!s^2) }\right\}
  \ , \\
\end{split}
\label{Ijacobi-eq-N1}
\end{equation}
and for the corresponding probability (\ref{Ijacobi-N2})
\begin{equation}
 \begin{split}
  E_2(0;I)
  & = {(1\!-\!s^2)^{2\alpha+2} s^4  \over
       4^{2\alpha+2}(2\alpha\!+\!3)B^2(\alpha\!+\!2,\alpha\!+\!2)}
      {}_{2}F_{1}(\alpha\!+\!\thalf,1;\alpha\!+\!2;1\!-\!s^2)
      {}_{2}F_{1}(\alpha\!+\!\fhalf,1;\alpha\!+\!2;1\!-\!s^2)
  \ , \\
  \sigma(s)
  & = (\alpha\!+\!1) \left\{
      { 1 \over s\, {}_{2}F_{1}(\alpha\!+\!\thalf,1;\alpha\!+\!2;1\!-\!s^2) }
    + { 1 \over s\, {}_{2}F_{1}(\alpha\!+\!\fhalf,1;\alpha\!+\!2;1\!-\!s^2) }
                     \right\}
  \ , \\
  G(s)
  & = (\alpha\!+\!2)s + 2(\alpha\!+\!1) \left\{
      { 1 \over s\, {}_{2}F_{1}(\alpha\!+\!\thalf,1;\alpha\!+\!2;1\!-\!s^2) }
    - { 1 \over s\, {}_{2}F_{1}(\alpha\!+\!\fhalf,1;\alpha\!+\!2;1\!-\!s^2) }
                     \right\}
  \ ,
\end{split}
\label{Ijacobi-eq-N2}
\end{equation}
in terms of the Gauss hypergeometric function $ {}_{2}F_{1}(a,b;c;z) $.
\end{proposition}
Proof - we proceed in a parallel manner as in the proof of Proposition 11
and the relations given there, however also using the linear transformation
formulae for the Hypergeometric functions. One such an example is
\begin{equation}
   4^{\alpha}B(\alpha\!+\!1,\alpha\!+\!1) 
     - s\, {}_{2}F_{1}(-\alpha,\half;\thalf;s^2) =
       \half{ s(1\!-\!s^2)^{\alpha+1} \over \alpha\!+\!1}
       {}_{2}F_{1}(\alpha\!+\!\thalf,1;\alpha\!+\!2;1\!-\!s^2) \ .
\end{equation}
We also require the comparable contiguous relation
\begin{equation}
   - (2\alpha\!+\!3)s^2 {}_{2}F_{1}(\alpha\!+\!\fhalf,1;\alpha\!+\!2;1\!-\!s^2)
   + {}_{2}F_{1}(\alpha\!+\!\thalf,1;\alpha\!+\!2;1\!-\!s^2)
   + 2(\alpha\!+\!1) = 0 \ .
\label{Icontig-reln}
\end{equation}
$ \square $

One can show that these two specific cases are solutions to our third order
differential equation (\ref{Ijacobi-ode3}), after noting the above contiguous
relation linking the two Hypergeometric functions for the case of $ N = 2 $.

The special case of $ \alpha = \beta = 0 $ and general $ N $ is no longer
a simple case as was the situation for the endpoint interval.

\subsection{Jacobi to Hermite Limit for the Interior Interval}
As we have shown in Subsection 4.3 one would also expect the limit 
$ \alpha \to \infty $ would recover the result for the Hermite ensemble
on $ (-t,t) $, under the scaling in (\ref{j2g-xfm}). This result is given
in \cite{TW-94}, although not in a form useful for our purposes. In that
reference a second order ODE is given for $ \tilde{R}(t) $ whereas we
want the corresponding ODE for $ R(t) $. Using the same kind of elimination
indicated in the proof of Proposition 4 on their Equations (5.31) and (5.32)
we find the result
\begin{equation}
  tr''+2r' = -2t[t-g] - 2g\sqrt{ (a')^2 + 4t[t-g]a - 2Nt^2[t-g]^2} \ ,
\label{gauss-ode2}
\end{equation}
where $ g = \sqrt{t^2+2r'} $ and $ a(t) = tr(t) $.
To verify that (\ref{Ijacobi-ode3}) has the correct limiting behaviour, we 
first find that the substitution (\ref{j2g-xfm}), to leading order in 
$ \alpha^2 $ yields
\begin{align}
 & r''' - 2 - {(r'')^2 \over r'} - {2\over t^{2}}r'
   + {2 \over t^{2}}{r^2 \over r'} g[t+g]
 \nonumber \\*
 & \qquad
    + 2r' g[t-g] - \left(4r-{2 \over t}\right)g
    + {(tr''-2r')^2 \over 2tr'g^2}[t-g] = 0
\label{Ij2g-ode3}
\end{align}
and that differentiation of the second order Hermite ODE, (\ref{gauss-ode2}),
gives
\begin{align}
 tr''' + 3r'' = 
 & -4t + 4g^2(tr'+r-2Nt) + 2g
 \nonumber \\*
 & - 8t(r-Nt)g + (t+r'')(2t^2+tr''+2r')/g^2
\label{Igauss-ode3}
\end{align}
with $ g $ as before.
By repeating the steps described in Subsection 4.3 we can show that both
(\ref{gauss-ode2}) and (\ref{Igauss-ode3}) can be combined to give the
limiting case, (\ref{Ij2g-ode3}).

\vfill\eject
\setcounter{equation}{0}

%%  wfc.tex
%%  New section (Section 5) of Witte-Forrester paper titled, 
%%  Gap probabilities for edge intervals in finite Gaussian and Jacobi 
%%       unitary matrix ensembles. 
%%  Replaces and supersedes former Section 6 (Appendix).  
%%  Contains solutions of equations: 
%%  (3.19):                          P-V 
%%  (3.20),(3.26),(4.49):            P-V 
%%  (4.50):                          P-V 
%%  (4.82),(4.83),(4.84):            P-V
%%  (4.29),(4.73):                   P-VI 
%%  A recent Tracy-Widom equation:   P-V. 
%%  (6.1),(6.6):  P-VI (not included---see Haine and Semengue)
\section {Reductions to Painlev\'e transcendents}

The analysis in the preceding sections yielded a number of nonlinear 
differential equations of the second and third order.  In this section 
we present their solutions in terms of the fifth and sixth Painlev\'e 
transcendents.  In all cases, we can arrive at one of the equations 
of the second order and second degree found by Chazy \cite{Ch1909b,Ch1911} 
and subsequently rederived by a number of authors.  

We begin with equation (\ref{gauss-ode-b}) for the variable $\tilde{R}(s)$.  
This equation is equivalent under a gauge transformation to an equation 
first derived by Chazy by transformation of the \mbox{Painlev\'e-V} 
equation.  It is the fourth member of the set of five equations denoted 
(II) in \cite{Ch1909b} and (C) in \cite[p.\,342]{Ch1911}.  It was 
also derived by Bureau \cite[pp.\,204--206]{B3} in his investigation 
of second-order second-degree equations.  

The solution of (\ref{gauss-ode-b}) is 
\begin{equation}
\tilde{R}(s) \,=\, \frac{\ep_1 w' - 2sw}{4w(w - 1)} \,-\, 
     \frac{N(w - 1) + \ep_1}{4sw}\,,  \eq{5.1} 
\end{equation}
where $\ep_1:= \pm 1$ and $w(s)$ (not to be confused with $ w $ used previously)
satisfies the differential equation,
\begin{align}
w'' &\,=\, \left\{ \frac{1}{2w} + \frac{1}{w-1} \right\} (w')^2 
     \,-\, \frac{1}{s}\,w'  \nonumber \\* 
&\qquad\quad\null \,+\, \frac{(w - 1)^2}{2s^2} 
     \left\{ N^2 w - \frac{(N - \ep_1)^2}{w} \right\} 
     \,+\, 2\ep_1 w \,-\, \frac{2s^2 w(w + 1)}{w - 1}\,.  \eq{5.2}
\end{align}
The prime denotes $d/ds$.  Under the change of variable, \mbox{$x = s^2$}, 
the latter equation becomes the standard \mbox{Painlev\'e-V} equation, 
\begin{align}
\frac{d^2 w}{dx^2} &\,=\, \left\{\frac{1}{2w} + \frac{1}{w - 1}\right\} 
     \left(\frac{dw}{dx}\right)^2 \,-\, \frac{1}{x}\,\frac{dw}{dx}  
     \nonumber \\*
&\qquad\quad\null \,+\, \frac{(w - 1)^2}{x^2} 
     \left\{\al w + \frac{\be}{w}\right\} \,+\, \frac{\ga w}{x} 
     \,+\, \frac{\de w(w + 1)}{w - 1}\,, \eq{5.3} 
\end{align}
with parameters, 
\begin{equation}
\al = \tfrac{1}{8}N^2, \qquad \be = - \tfrac{1}{8}(N - \ep_1)^2, \qquad 
     \ga = \half\ep_1, \qquad \de = - \half\,.  \eq{5.4} 
\end{equation}

In terms of the same $w(s)$, the solution of the companion equation 
(\ref{gauss-ode}) is 
\begin{align}
R(s) &\,=\, \frac{1}{8sw(w - 1)^2}\, 
     \Bigl\{ sw' + N(w - 1)^2 + (2s^2 - 1)w + 1 \Bigr\}  \nonumber \\*
&\qquad\quad\null \times 
     \Bigl\{ sw' - N(w - 1)^2 - (2s^2 + 1)w + 1 \Bigr\},  \eq{5.5}
\end{align}
and the auxiliary variable $h(s)$ is given by 
\begin{equation}
h(s) \,=\, \frac{2sw^2 - \ep_1 w'}{2w(w - 1)} 
     \,+\, \frac{N(w - 1) + \ep_1}{2sw}\,.  \eq{5.6}
\end{equation}

The last two equations also furnish the solution of (\ref{gauss-ode-a}) with 
$h(s)$ being identified with the square root with the upper sign.  The identity 
(\ref{gauss-sqr}) is satisfied identically by the forms of $R(s)$ and 
$\tilde{R}(s)$ given here.  They also solve equation (\ref{j2g-ode3}) after 
renaming the variables in (\ref{j2g-ode3}) according to \mbox{$t \to s$}, 
\mbox{$r(t) \to R(s)$}.  
In that case, the parameter $N$ plays the role of the third integration 
constant because (\ref{j2g-ode3}) can be obtained from (\ref{gauss-ode}) by 
differentiating out the parameter~$N$.  

On the other hand, equation (\ref{gauss-ode3}), which was also obtained by 
differentiation of (\ref{gauss-ode}), is a true generalisation of 
(\ref{gauss-ode}).  Its first integral is 
\begin{align}
sR'' + 2R' &\,=\, 2s(s - h)  \nonumber \\*
&\qquad\null \,-\, 2h \sqrt{(R + sR')^2 - 4s^2(s - h)R 
     - 2Ns^2(s - h)^2 + K_1}\,,  \eq{5.7}
\end{align}
where $h$ denotes $\sqrt{s^2 - 2R'}\,$ and $K_1$ is the integration 
constant.  Although the case of nonzero $K_1$ is not relevant 
to the discussion in Section~4, we nevertheless have another integrable 
equation.  Its solution is 
\begin{align}
R &\,=\, \frac{1}{8sw(w - 1)^2} 
     \Bigl\{ sw' + N(w - 1)^2 + (2s^2 - 1)w + 1 \Bigr\}  \nonumber \\*
&\qquad\quad\null \times 
     \Bigl\{ sw' - N(w - 1)^2 - (2s^2 + 1)w + 1 \Bigr\}  \nonumber \\*
&\qquad\quad\null \,-\, 
     \frac{K(w + 1)\bigl\{ N(w - 1)^2 - 2s^2 w \bigr\}}{2sw(w-1)}  
     \nonumber \\*
&\qquad\quad\null \,+\, \frac{K^2(w - 1)(3w + 1)}{2sw}\,,  \eq{5.8} \\
h &\,=\, \frac{2sw^2 - \ep_1 w'}{2w(w-1)} \,+\, 
     \frac{(N - 2K)(w - 1) + \ep_1}{2sw}\,,  \eq{5.9}
\end{align}
where $\ep_1:= \pm 1$,\, \mbox{$K_1 = 8K^2(N - 2K)$}, and $w(s)$ 
satisfies the differential equation, 
\begin{align}
w'' &\,=\, \left\{ \frac{1}{2w} + \frac{1}{w-1} \right\} (w')^2 
     \,-\, \frac{1}{s}\,w'  \nonumber \\* 
&\qquad\quad\null \,+\, \frac{(w - 1)^2}{2s^2} 
     \left\{ (N - 2K)(N + 6K) w \,-\, \frac{(N - 2K - \ep_1)^2}{w} 
     \right\}  \nonumber \\*
&\qquad\quad\null \,+\, 2(4K + \ep_1) w 
     \,-\, \frac{2s^2 w(w + 1)}{w - 1}\,.  \eq{5.10}
\end{align}
As before, the change of variable \mbox{$x = s^2$} transforms (\ref{5.10}) 
to a standard \mbox{Painlev\'e-V} equation.  When $K = 0$, these results 
reduce to the above results for (\ref{gauss-ode}).  By changing the sign of~$R$,
these formulae, with zero or nonzero $K_1$ as appropriate, also 
solve equations (\ref{gauss-ode2})--(\ref{Igauss-ode3}).  
%%  Check whether (\ref{gauss-ode3}) is fully integrable.  YES IT IS!

We now turn our attention to equation (\ref{Ejacobi-ode3}), which presents a 
somewhat greater challenge.  This differential equation is of the third order 
and second degree.  We find that it is more manageable when written in 
terms of the auxiliary variable $H(s)$ rather than~$\sigma(s)$.  After 
that, we observe a slight improvement by choosing a new variable $y(s)$, 
in terms of which we have 
\begin{align}
H(s) &\,=\, (1 - s^2)y - \al_1 s,  \eq{5.11}  \\
\sigma(s) &\,=\, 
     \tfrac{1}{4} \Bigl\{ \al_1 s^2 y' - (s^2 - 1)y^2 \Bigr\}^{-1} 
     \Bigl\{ s^2(s^2 - 1)^2 \bigl( yy''' - y'y'' \bigr)  \nonumber \\*
&\qquad\quad\null + 2s(s^2 - 1) \bigl[ (4s^2 - 1)yy'' - (2s^2 - 1)(y')^2 
     \bigr]  \nonumber \\*
&\qquad\quad\null - 2y \bigl[ 2s^2(s^2 - 1)^2 y^2 + 3\al_1 s^3(s^2 - 1)y 
     - 6s^4 + 3s^2 + 1 \bigr] y'  \nonumber \\*
&\qquad\quad\null - 2sy^2 \bigl[ (s^2 - 1)(3s^2 - 1)y^2 
     + 2\al_1 s(3s^2 - 2)y  \nonumber \\*
&\qquad\quad\null + 2(\al_1^{\,2} - 1)s^2 \bigr] \Bigr\}.  
     \eq{5.12}
\end{align}
The differential equation satisfied by $y(s)$ takes the form, 
\begin{align}
\bigl\{ 2\sigma - \al_1 s^2 y \bigr\}^2 &\,=\, 
     \null - s^2(s^2 - 1)^2\bigl\{ yy'' - (y')^2 \bigr\}  \nonumber \\*
&\qquad\quad\null - 2s^3(s^2 - 1)yy' + y^2 \bigl\{ s^2(s^2 - 1)^2 y^2  
     \nonumber \\*
&\qquad\quad\null + 2\al_1 s^3(s^2 - 1)y + (\al_1^{\,2} - 1)s^4 + 1 
     \bigr\},  \eq{5.13}
\end{align}
where it is understood that $\sigma$ is to be replaced by the right-hand 
side of~(\ref{5.12}).  This differential equation of the third order 
and second degree admits the first integral, 
\begin{align}
&\Bigl\{ s(s^2 - 1)^2 y'' + 2(s^2 - 1)(2s^2 - 1)y' \nonumber \\*
&\qquad\null + 8s^3y^3 + 12\al_1 s^2 y^2 + s(2s^2 + K_1 - 6)y \Bigr\}^2 
     \nonumber \\*
&\qquad\qquad\quad \,=\, 4 \Bigl\{ (s^2 + 1)y + \al_1 s \Bigr\}^2 
     \Bigl\{ (s^2 - 1)^2 (sy' + y)^2  \nonumber \\*
&\qquad\qquad\qquad\qquad\null + s^2y^2 \bigl( 
     4s^2y^2 + 8\al_1 sy + K_1 - 4 \bigr) \Bigr\},  \eq{5.14}
\end{align}
where $K_1$ is the constant of integration.  

Equation (\ref{5.14}) is equivalent under a gauge transformation to 
the fifth member of the aforementioned set of Chazy equations.  
Chazy found this equation by transforming the \mbox{Painlev\'e-VI} 
equation, but he did not show the actual formula.  Bureau also obtained 
this equation \cite[pp.\,200--202]{B3} but did not solve it.  The 
same equation in a different gauge was obtained by Fokas and Yortsos 
\cite{FY}, who gave the reduction to \mbox{Painlev\'e-VI}.  

The variables $y(s)$, $H(s)$, and $\sigma(s)$ are given in terms of 
a function $w(s)$ by the formulae, 
\begin{align}
y &\,=\, \frac{\ep_1 s(s^2 - 1)w' 
     - (w - 1)\bigl\{ \ep_1(w + s^2) + 2\al_1 s^2 \bigr\}}
     {2s(s^2 - 1)w}\,,  \eq{5.15} \\
H &\,=\, \frac{\ep_1 s(1 - s^2)w' + \ep_1 (w - 1)(w + s^2) 
     - 2\al_1 s^2}{2sw}\,,  \eq{5.16} \\
\sigma &\,=\, \null - \frac{\bigl\{ s(s^2 - 1)w' - (w - 1)(w + s^2) 
     \bigr\}^2} {8sw(w - 1)(w - s^2)}  \nonumber \\*
&\qquad\qquad\null \,-\, \frac{s(w - 1)\bigl\{ (K_1 - 4)w 
     - 4\al_1^{\,2}s^2 \bigr\}} {8w(w - s^2)}\,,  \eq{5.17}
\end{align}
where \mbox{$\ep_1 := \pm 1$}.  The function $w(s)$ satisfies the 
differential equation, 
\begin{align}
w'' &\,=\, \frac{1}{2}\left\{ \frac{1}{w} + \frac{1}{w-1} 
     + \frac{1}{w - s^2} \right\} (w')^2  \nonumber \\*
&\qquad\quad\null \,-\, \left\{ \frac{1}{s} + \frac{2s}{s^2 - 1} 
     + \frac{2s}{w - s^2} \right\} w'  \nonumber \\*
&\qquad\quad\null \,+\, \frac{w(w - 1)(w - s^2)}{2s^2(s^2 - 1)^2}  
     \nonumber \\*
&\qquad\quad\null \times \biggl\{ 1 - \frac{(1 + 2\ep_1\al_1)^2 s^2}{w^2} 
     + \frac{(K_1 - 4\al_1^{\,2})s^2(s^2 - 1)}{(w - s^2)^2} \biggr\}. 
     \eq{5.18}
\end{align}
Under the change of variable, \mbox{$x = s^2$}, the latter equation 
becomes the \mbox{Painlev\'e-VI} equation, 
\begin{align}
\frac{d^2 w}{dx^2} &\,=\, \frac{1}{2}\left\{ \frac{1}{w} + \frac{1}{w-1} 
     + \frac{1}{w-x} \right\} \left(\frac{dw}{dx}\right)^2 \nonumber \\*
&\qquad\quad\null \,-\, \left\{ \frac{1}{x} + \frac{1}{x-1} 
     + \frac{1}{w-x} \right\}\frac{dw}{dx}  \nonumber \\*
&\qquad\quad\null \,+\, \frac{w(w-1)(w-x)}{x^2(x-1)^2} \biggl\{ \al 
     + \frac{\be x}{w^2} + \frac{\ga(x-1)}{(w-1)^2} 
     + \frac{\de x(x-1)}{(w-x)^2} \biggr\},  \eq{5.19}
\end{align}
with parameters,
\begin{equation}
\al = \tfrac{1}{8}\,, \qquad 
     \be = \null - \tfrac{1}{8}(1 + 2\ep_1\al_1)^2, \qquad 
     \ga = 0, \qquad 
     \de = \tfrac{1}{8}(K_1 - 4\al_1^{\,2}).  \eq{5.20}
\end{equation}
These formulae also solve (\ref{Ijacobi-ode3}) with the sign of~$\sigma$ 
changed.  

We draw the reader's attention to another set of five second-order 
second-degree Chazy equations which make regular appearances in 
random matrix theory.  This set is denoted (III) in \cite{Ch1909b} 
and (B) in \cite[p.\,340]{Ch1911} and has also been studied by 
Bureau \cite{B2,B3}, Fokas and Ablowitz \cite{FA1}, Jimbo and Miwa 
\cite[Appendix~C]{JM2}, and Cosgrove and Scoufis~\cite{SD}.  This set 
can be embraced in a single equation, 
\begin{align}
(y'')^2 &\,=\, \null - \frac 4{g^2(x)} \Bigl\{ c_1(xy' - y)^3 
     + c_2y'(xy' - y)^2 + c_3(y')^2(xy' - y)  \nonumber \\*
&\qquad\quad\null + c_4(y')^3 + c_5(xy' - y)^2 + c_6y'(xy' -  y) 
     + c_7(y')^2  \nonumber \\*
&\qquad\quad\null + c_8(xy' - y) + c_9y' + c_{10} \Bigr\},  \eq{5.21}
\end{align}
where $g(x) := c_1 x^3 + c_2 x^2 + c_3 x + c_4$ and the prime denotes 
$d/dx$.  This is equation \mbox{SD-I}, introduced in 
\cite[pp.\,57,\,65--73]{SD}.  It was called there the ``master Painlev\'e 
equation'' because it unified all of the six Painlev\'e transcendents 
into a single equation.  Only four of the $10$ parameters in \mbox{SD-I} 
are essential because the equation retains its shape under gauge 
transformations of the form,
\begin{equation}
\bar{x} \,=\, \frac{ax + b}{cx + d}\,, \qquad 
     \bar{y} \,=\, \frac{hy + kx + m}{cx + d}.  \eq{5.22}
\end{equation}
By using this gauge freedom appropriately, equation \mbox{SD-I} 
can be split into six normalised forms, whose solutions in terms 
of Painlev\'e transcendents are given in~\cite{SD}.  

The full version of \mbox{SD-I} generates an abundant supply of 
third-order equations under differentiation.  For example, replacing 
$c_i$ by \mbox{$c_i + Kc_{i+6}$} for $i = 5$, $\dots\,$,~$10$ and 
differentiating out the parameter $K$ produces a $15$-parameter 
equation (after normalising $c_{11}$, say, to $1$ or~$0$).  In a 
similar fashion, one can generate a $25$-parameter equation of the 
third order and second degree.  These big equations are all integrable 
in terms of \mbox{Painlev\'e-VI} in the generic case and one of the 
other five Painlev\'e transcendents or elliptic functions in the 
remaining cases.  

A special case of the aforementioned $15$-parameter equation 
appears, for example, in Tracy and Widom~\cite{TW-94}.  This case 
was solved recently in terms of \mbox{Painlev\'e-VI} by Haine and 
Semengue~\cite{HS-99}.  A more recent example, also by Tracy and 
Widom~\cite{TW1999}, is the equation, 
\begin{align}
y''' &\,=\, \frac{1}{2} \left\{ \frac{1}{y'} + \frac{1}{y' - 1} 
     \right\} (y'')^2 \,-\, \frac{1}{x}\,y''  \nonumber \\*
&\qquad\quad\null \,-\, \frac{2(k + n)}{x}\,y'(y' - 1) \,+\, 
     \frac{x + n}{2x^2}\,\bigl( n - x + 2y \bigr)  \nonumber \\*
&\qquad\quad\null \,-\, \frac{(n + y)^2}{2x^2 y'} 
     \,-\, \frac{(x - y)^2}{2x^2(y' - 1)}\,,  \eq{5.23} 
\end{align}
the prime denoting $d/dx$.  This equation has the first integral, 
\begin{align}
x^2(y'')^2 &\,=\, \null - 4(k + n)x(y')^3  \nonumber \\*
&\qquad\quad\null \,+\, \bigl\{ 4(k + n)y + x^2 
     + 2(2k + 3n)x \bigr\}(y')^2  \nonumber \\*
&\qquad\quad\null \,-\, \bigl\{ 2(x + 2k + 3n)y + 2nx + n^2 
     \bigr\}y'  \nonumber \\*
&\qquad\quad\null \,+\, (y + n)^2 \,-\, 4K_1 y'(y' - 1),  \eq{5.24}
\end{align}
where $K_1$ is the constant of integration.  

The latter equation is gauge-equivalent to the normalised form 
\mbox{SD-I.b} in~\cite{SD}.  Its solution is 
\begin{align}
y &\,=\, \frac{1}{4(k + n)w} \left\{ \frac{xw'}{w - 1} - w \right\}^2 
     \nonumber \\*
&\qquad\quad\null \,-\, \frac{(w - 1)\bigl\{ k(w - 1) + n(w + 1) 
     \bigr\}}{4w}  \nonumber \\*
&\qquad\quad\null \,+\, \frac{1}{4(k + n)} 
     \biggl\{ \frac{2nx(w - 2)}{w - 1} \,-\, \frac{x^2 w}{(w - 1)^2} 
     \,+\, 2kx \,+\, \frac{4K_1}{w} \biggr\},  \eq{5.25}
\end{align}
where $w(x)$ satisfies the \mbox{Painlev\'e-V} equation (\ref{5.3}) 
with parameters, 
\begin{equation}
\al = \half (k + n + \ep_1)^2, \qquad 
     \be = \half (n^2 - k^2 + 4K_1), \qquad 
     \ga = - n, \qquad \de = - \half\,,  \eq{5.26}
\end{equation}
with $\ep_1:= \pm 1$.  

\vfill\eject

\bibliographystyle{siam}
\bibliography{moment,random_matrices,nonlinear}
\vspace*{30mm}
\begin{center}
\bf Acknowledgements
\end{center}
One of the authors (NSW) would like to acknowledge the support of a
Australian Research Council large Grant whilst this work was performed.
\vfill\eject

\end{document}